\documentclass{JHEP3}

\usepackage{epsfig,multicol}

\newcommand{\no}{\nonumber\\}
\newcommand{\be}{\begin{equation}}
\newcommand{\ee}{\end{equation}}
\newcommand{\ba}{\begin{eqnarray}}
\newcommand{\ea}{\end{eqnarray}}
\newcommand{\ci}[1]{\cite{#1}}
\newcommand{\bi}[1]{\bibitem{#1}}
\newcommand{\la}[1]{\label{#1}}
\def\gl#1{(\ref{#1})}

\def\na{\!\not\!\nabla}
\def\di{\not\!\!D}
\def\ddi{\not\!\!{\bf D}}

\def\O{\rm O}
\def\ma{>}
\def\mi{<}

\newcommand\fverb{\setbox\pippobox=\hbox\bgroup\verb}
\newcommand\fverbdo{\egroup\medskip\noindent%
                        \fbox{\unhbox\pippobox}\ }
\newcommand\fverbit{\egroup\item[\fbox{\unhbox\pippobox}]}
\newbox\pippobox

\title{Brane world generation by matter and gravity}

\author{A.A. Andrianov $^{\flat\sharp}$, 
V.A. Andrianov $^{\flat}$, 
P. Giacconi $^{\sharp}$ and 
R. 
Soldati$^{\sharp}$\\
$^{\flat}$ V.A.Fock Department of Theoretical Physics,
Sankt-Petersburg State University,\\
198504 Sankt-Petersburg, Russia\\
$^{\sharp}$ Dipartimento di Fisica, 
Universit\'a
di Bologna and\\
 Istituto Nazionale di Fisica Nucleare, Sezione di 
Bologna,\\
40126 Bologna, Italia}

\abstract{We present a non-compact {\it 4 + 1}
dimensional model with a local strong four-fermion interaction
supplementing it with gravity. In the strong
coupling regime it reveals the spontaneous translational symmetry
breaking which eventually leads to the formation  of domain walls, 
or thick 3-branes, embedded in the AdS$_5$ manifold. To describe this
phenomenon we construct the appropriate low-energy effective Action 
and find kink-like vacuum solutions in the quasi-flat Riemannian
metric. We discuss the generation of ultra-low-energy {\it 3 + 1} 
dimensional physics and we establish the relation among the bulk five
dimensional gravitational constant, the brane Newton's constants and 
the curvature of AdS$_5$ space-time. The plausible relation between 
the compositeness scale of the scalar matter and the symmetry breaking 
scale is shown to support the essential decoupling of branons, 
the scalar fluctuations of the brane, from the Standard Model matter, 
supporting their possible role in the dark matter saturation.
The induced cosmological constant on the brane does vanish due to exact 
cancellation of matter and gravity contributions.}


\preprint{hep-th/0503115}

\dedicated{\large Dedicato ad Alessandra Tallei}

\begin{document}

\section{Introduction}
The conjecture about that our {\it 3 + 1} dimensional world might be
allocated on a brane (or domain wall) in a multi-dimensional space-time  
has recently invoked much activity \ci{R-S}-\ci{RS1}, owing to the new 
tools it provides to solve the long standing mass and scale hierarchy  
problems \ci{ADD}-\cite{fmasshi2} in particle theory. 
New extra dimensional physics could manifest itself in accessible
experiments and observations, when the size of extra dimensions is
relatively large \ci{ADD,ADD2} or even infinite
\ci{R-S}, \ci{infin1} -- \ci{infin4}, 
in as much to feed 
research programs in running and forthcoming collider and non-collider 
experiments 
\ci{ADD2}, \ci{KKgrav1}--\ci{missch}, \ci{bran3}, \ci{fluct}.
Nowadays the brane world scenarios and their applications are well 
summarized in a dozen of review articles
\ci{rev1}--\ci{rev10}.

On the one hand, the brane itself is often considered as an 
{\sl elementary} geometrical object of a vanishing thickness along
the extra dimensions, as it is promoted by  superstring
theories (with supersymmetric compactifications, see \ci{rev2} and references therein,
and non-supersymmetric ones, see \ci{koko} and references therein). 
In other words, this kind of brane represents just a
boundary for higher dimensional objects. On the other hand, such an approach
gives rise to the question about the origin of branes trapping our
matter world. In addition to, when endowing branes with a tiny thickness,
one allows some more options for a solution of the mass hierarchy
problem \ci{fmasshi1,fmasshi2,masshi1,masshi2} (see also reviews \ci{rev1,rev3,rev9}).

A non-antagonistic alternative to the infinitely thin branes is 
provided by an effective multi-dimensional field theory, 
owing to the possibility of a spontaneous breaking of the
translational symmetry. The thick (or fat) brane (or domain wall) 
formation and the trapping of light particles
in its layer might be obtained \ci{scal1}--\ci{galoc3} by a number of particular background
scalar and/or gravitational fields living in the multi-dimensional bulk,
when their vacuum configuration has a non-trivial topology, 
thereby generating zero-energy states localized on the brane.

Respectively, the mechanism of how such background fields might emerge and 
further induce the spontaneous breaking of translational symmetry is worthy 
to be elaborated and the  domain wall creation, due to the self-interaction
of certain particles in the bulk, may become a conceivable and appealing 
possibility \ci{aags}.

In this paper we continue the exploration of a non-compact
{\it 4 + 1} dimensional fermion model \ci{aags} with a local strong 
four-fermion interaction, by supplementing it with a partially induced 
background gravitational field. Both kinds of interactions will lead coherently 
first to the discrete symmetry breaking and, further on, to the breaking of translational 
invariance. This can be achieved in terms of a particular domain wall pattern of the 
vacuum state \ci{Rajar}-- \ci{Baz}, 
just allowing the light massive Dirac particles to live essentially in 
{\it 3 + 1} dimensions. Those very same interactions generate localized zero-modes for the
composite scalar fields, just completing the fermion matter content on the brane 
with a scalar counterpart.

We shall concentrate ourselves upon 
the main dynamical origin of the spontaneous symmetry breaking
-- on fermion self-interaction supplemented by gravity -- and yet simplify
the model by neglecting all gauge field
interaction. In this sense, our model may be considered a sector
of the so called {\sl domain wall gravitating standard model}.

The five dimensional fermion model with spin-0 and spin-2 induced
self-interaction is formulated in Section 2. There the fermion
self-interaction {\it via} the spin-0 channel is 
restricted to a four-fermion type that will be sufficient to
trigger the localization of massive Dirac fermions. It contains two
dimensional coupling constants expressed in units of the compositeness
scale $\Lambda$. The latter one plays the role of a cut-off for virtual
fermion energies and momenta. 

Meanwhile, the interaction {\it via} the spin-2 channel ({\sl extra dimensional gravity}) 
is introduced in a non-linear way, in order to make this model covariant under the 
space-time diffeomorphisms, with a specific five dimensional Einstein-Hilbert bare Action
and a bare cosmological constant to balance the formation
of the physical Newton's and cosmological constants on the brane. 
Among different options, we pay  attention to the case when the
five dimensional gravity is fundamental and its related
Einstein-Hilbert term is not too much corrected by fermion induced
effective Action. As well, we shall turn ourselves 
to the economical scenario of induced gravity when, contrary to the
previous case,  the bare gravitational constant may be taken negligible 
and gravity is principally induced by fermion matter being therefore composite.

In this Section 2, once the
scalar bosonization of the four-fermion interaction has been
implemented, the low-energy effective Action for composite scalar and
gravity fields is obtained in the mean field or large-$N$ approximation, 
where $N$ roughly counts the number of fermion species in the Standard Model. 
This effective Action, which arises out of fermion one-loop radiative
corrections, does accumulate the radiative contributions of high-energy
fermion virtualities to describe infrared phenomena of spontaneous
symmetry breaking. It already contains the kinetic terms for the scalar
auxiliary fields, endowing them with the structure of composite fields. 
In the calculation of that effective Action, the Euclidean space-time
approach for the invariant cut-off and the finite-mode regularization \ci{AnBo}
for separation of high- and low-energy fermion fields are employed.

In Section 3 we search for classical vacuum configurations of 
gravity and scalar fields by analyzing the low-energy
effective Action. This search is restricted to the class of conformal-like
metrics (warped geometries) with the flat Minkowski hyperplanes at
each point along the fifth coordinate. The joint solution of
Einstein and non-linear Klein-Fock-Gordon equations is properly found within the 
weak-gravity approximation, {\it i.e.} assuming a relatively small 
five dimensional gravitational constant, what will be eventually justified
in Section 4 after normalization to the Newton's gravitational constant on
the brane. As expected to the leading order, the equations of
motion for the scalar fields are not affected by gravity and their solutions
coincide with the flat case investigated in \ci{aags}. In this sense,
in the model under discussion, the brane and warped geometry creation
is basically maintained by matter but not by gravity\footnote{The interplay
between
gravity and matter effects in brane world formation and corresponding
Einstein equations have been  subjects
of extensive studies in different models 
\ci{R-S,rev1,rev5,rev10,scal1,scal2,scal3,scal5,gravbr}.}. 
 
We select out the magnitudes of our four-fermion coupling constants
in such a way that the dynamical breaking of the so called $\tau$-symmetry
and translational invariance were supported and, moreover, that the  
brane located Dirac fermions were supplied  with masses. 
The characteristic scale $M$ of
the symmetry breaking -- inverse thickness of a brane -- can be
conceivably much less than the compositeness scale $\Lambda\,,$ keeping
consistently the brane formation and particle localization phenomena
in the low-energy region.

The remaining pair of the Einstein-like field equations can be 
rearranged so that one of them does specify the warp factor of the
five dimensional anti-de Sitter (AdS$_5$) geometry, whereas the other 
one does represent the integral of motion with the bulk cosmological 
constant as an integration constant.
The latter property holds exactly in all order of gravitational 
perturbation theory. As a consequence, this integration constant 
is found to be firmly fixed by the vacuum (kink-like) configurations 
of the scalar fields and of the warped geometry.

In Section 4 we examine the spectral properties of light particles
allocated on the brane, we describe the structure of the ultra-low energy
Lagrange density and finally summarize the results of \ci{aags} for  
the coupling constants of light particle interactions, 
which are parameterized by the ratio $M/\Lambda$ of 
the symmetry breaking and the compositeness scales. 
The mass spectrum of light particle together with their interactions are described 
up to the leading order in the gravitational constant. 
We argue that this leading contribution into the brane world matter interaction is, 
in fact, independent of the five dimensional gravity, in spite of the
nonperturbative quantum tunneling of massive states off the brane
\ci{galoc2} in the AdS$_5$ geometry.     

Section 5 is devoted to the estimation of all the scales and the
coupling constants we have previously introduced. This can be achieved
by imposing the observed value of the  Newton's constant and
the Newton's gravitational law within the presently available limits
in modern experiments \ci{adelb}. It is adopted that 
both the scale of compositeness and the
translational symmetry breaking scale must be naturally very high,
in between several TeV's and the GUT scale $10^{15}$ GeV. 
Therefrom it is derived that, if five dimensional gravity is
fundamental and displaying some Planck-like scale much larger 
than the compositeness scale, then there is a window for a non-trivial 
(but weak) interaction between brane fluctuations (branons \ci{bran3,bran1,bran2}), 
Higgs-like scalars and fermions, with a low compositeness scale of the order
tens of TeV's, the trapping barrier for light particles as high as $2\div 3$
TeV and the AdS$_5$ curvature scale $\sim 10^{-3}$ eV, which is comparable with the 
existing experimental checks of the Newton's law \ci{adelb}. 
 
On the other hand, if the induced gravity is dominated
at ultra-low energies, then the scalar matter interaction is highly
suppressed, just making branons good sterile candidates
to contribute into dark matter \ci{bran3}.  
The cosmological constant induced on the
brane is also calculated and found to vanish exactly, 
just making consistently endorsed the {\it ansatz} for the flat 
Minkowski's hyperplanes\footnote{This
  phenomenon of exact compensation between gravity and matter
  contributions into the brane cosmological constant has been also found  in
 \ci{R-S}, \ci{cosm1} -- \ci{cosm3} in
different brane-world models.}.

In the concluding Section we mainly summarize our observations on the interplay between
the five dimensional Planck-like mass, the compositeness and the dynamical symmetry
breaking scales and the AdS$_5$ curvature. As a further development,
the possibility of implementing an occasional tiny matter and vacuum energy defect,
located in the forth space direction in the presence of gravity, is
shortly discussed. As a matter of fact, in our previous paper \cite{aags} we
have demonstrated that such a defect may trigger and thereby justify 
the dynamical breaking of translational invariance at a particular place in 
the extra 
dimensions\footnote{Another realization of thin defects in extra
  dimensional world see in \ci{def}.} 
and it might supply the branon degrees of freedom with a small mass. 


\section{Five dimensional fermion model with scalar and gravity
induced self-interaction}

Let us remind \ci{aags} the domain wall phenomenology
 and introduce the necessary notations. We start from the model of
one four-component fermion bi-spinor field $\psi(X)$
defined on a five dimensional flat Minkowski space-time and
coupled to a scalar field $\Phi(X)$.
The extra-dimension coordinate is assumed to be space-like,
$$
X^A = (x^\mu, z)\ , \quad x^\mu = (x^0, x^1, x^2, x^3)\ , \quad
(\eta^{AA}) = (+,-,-,-,-)
$$
and the subspace of coordinates $x^\mu$ eventually corresponds to the
four dimensional Minkowski space.
The extra-dimension size is supposed to be infinite (or large enough).
The fermion wave function is then described by the Dirac equation
\be
[\,i\gamma^A \partial_A - \Phi(X)\,]\psi(X) = 0\ , \quad
\gamma^A = (\gamma^\mu, i\gamma_5)\ ,\quad \{\gamma^A, \gamma^B\}
= 2\eta^{AB}\ , \la{5dir}
\ee
$\gamma^\mu$ being a standard set of four dimensional Dirac matrices in the
chiral (or Weyl) representation with
$\gamma_5\equiv i\gamma^0\gamma^1\gamma^2\gamma^3$.

The trapping of light fermions on a four dimensional hyper-plane
-- the domain wall -- localized in the
fifth dimension at $z = z_0$ can be promoted
by a certain {\sl topological}, $z$-dependent configuration of
the vacuum expectation value of the scalar field  
\mbox{$\langle\Phi(X)\rangle_0 = \varphi (z)$} -- for
instance $ \varphi (z) =
M\, \mbox{\rm tanh}(Mz)$ --
due to the appearance of zero-modes with a certain chiralities in the spectrum
of the four dimensional Dirac operator \ci{R-S,rev1}.

If we aim to build up a light Dirac fermion, we need two different
chiralities for the same shape of scalar background. Then the minimal
set of fermions over the five dimensional space-time
has to include \ci{galoc2,aags} two proto-fermions $\psi_1(X), \psi_2(X)$.
In order to generate left- and right-handed
parts of a four dimensional Dirac bi-spinor as zero modes,
those fermions have to couple to the  scalar field $\Phi(X)$
with opposite charges,
\be
[\,i \not\!\partial - \tau_3\Phi(X)\,]\Psi(X) = 0\ ,\quad
\not\!\partial \equiv \widehat\gamma^A \partial_A\ ,\quad
\Psi(X) =\left\lgroup\begin{array}{c}\psi_1(X)\\
\psi_2(X)\end{array}\right\rgroup\ ,
\la{2fer}
\ee
where $\widehat\gamma^A \equiv \gamma^A\otimes {\bf 1}_2$ are
Dirac matrices and
$\tau_a \equiv {\bf 1}_4 \otimes \sigma_a,\ a=1,2,3 $
are the generalizations of
the Pauli matrices $\sigma_a$ acting on the bi-spinor components $\psi_i(X)$.

In addition to the trapping scalar field,
a further one is required to supply light domain wall fermions
with a mass. Its coupling must mix left and right chiralities as the
mass term breaks the chiral invariance.
Thus we introduce two types of four-fermion self-interactions
to reveal two composite scalar fields with  a proper coupling to fermions.
These two scalar fields
acquire mass spectra similar to fermions with light counterparts located
on the domain wall. The dynamical scheme of creation of domain wall
particles turns out to be quite economical and few predictions on masses
and decay constants of fermion and boson particles have been derived \ci{aags}.
However the allocation of matter on the domain wall certainly lead to
strong gravitational effects. Moreover, the gravity itself may cause
in turn the localization of the matter fields on a domain wall.

In the present paper we shall extend the dynamical mechanism of the fermion
self-interaction by including, partially or completely, 
the gravitational contribution as induced by  the high-energy spinor matter. 
Our main interest is focused on the
thick brane formation corresponding to the flat {\it 3+1} dimensional
Minkowski space, whereas the gravity is non-trivial in the
fifth direction orthogonal to the Minkowski's world. It is treated
self-consistently together with the vacuum configurations of composite
scalar fields.

Let us now formulate the fermion model in five 
dimensions\footnote{We extend the idea of the Top-mode Standard Model (TmSM)
\ci{Topmode} on extra dimensions. Other extensions of TmSM have been
undertaken in \ci{6top1,6top2}.}, which
implements the mechanism of translational symmetry breaking to
create a domain wall. It is described by the classical Lagrange density
\be
{\cal L}^{(5)} (\overline{\Psi},\Psi) = \overline{\Psi}\ i \!\not\!\partial \Psi +
\frac{g_1}{4N \Lambda^3}\left(\overline{\Psi}\tau_3 \Psi\right)^2
+ \frac{g_2}{4N \Lambda^3}\left(\overline{\Psi} \tau_1\Psi\right)^2,
\la{mod}
\ee
where
$\Psi(X)$ is an eight-component five dimensional
fermion field, see eq.\,\gl{2fer} --
either a bi-spinor in a four dimensional theory or a spinor in a
six-dimensional theory --
which may also realize a flavor and color multiplet with the total number
$N=N_f\,N_c$ of spinor degrees of freedom. We can say that, {\it grosso modo}, 
the number of color and flavor
degrees of freedom of massive fermions in the Standard Model is around twenty,
if all of them are originated from corresponding five dimensional proto-fermions. 
The ultraviolet cut-off scale $\Lambda$ bounds fermion momenta, 
as the four-fermion interaction is supposed here to be an effective one,
whereas $g_1$ and $g_2$ are suitable dimensionless
and eventually scale dependent effective couplings.

This Lagrange density can be more conveniently represented with the help of a pair of
auxiliary scalar fields $\Phi(X)$ and $H(X)$, which eventually will allow to
trap a light fermion on the domain wall and to supply it with a mass: namely,
\be
{\cal L}^{(5)} (\overline{\Psi},\Psi,\Phi, H) =
\overline{\Psi} ( i\!\not\!\partial
- \tau_3 \Phi -\tau_1 H) \Psi - \frac{N \Lambda^3}{g_1}\,\Phi^2
- \frac{N \Lambda^3}{g_2}\,H^2\ . \la{aux}
\ee
For sufficiently strong couplings,
this system undergoes
the phase transition to the state  in which the condensation of
fermion-antifermion pairs does spontaneously  break --
partially or completely -- the so-called $\tau$-symmetry: \mbox{$
\Psi \longrightarrow \tau_1 \Psi;\
\Phi \longrightarrow -\,\Phi$;} and $\Psi \longrightarrow \tau_3 \Psi;\
H \longrightarrow -\,H$.

In order to develop the infrared phenomenon of the $\tau$-symmetry breaking,
the effective Lagrange density containing  essential low-energy
degrees of freedom has to be suitably derived.
To this concern, we proceed
in the transition to the Euclidean space,
where the invariant momentum cut-off can be unambiguously implemented.
Within this framework, the notion
of low-energy is referred to momenta $|p| < \Lambda_0$ as compared with
the cut-off $\Lambda_0 \ll \Lambda$. However, after the elaboration
of the domain wall vacuum, we will search for the fermion states
with masses $m_f$ much lighter than the dynamical scale $\Lambda_0$,
{\it i.e.} for the ultralow-energy physics.
Thus, eventually, there are three scales in the present model
in order to implement the domain wall particle trapping.

Let us decompose the momentum space fermion fields into their high-energy part
$\Psi_h(p)\equiv\Psi(p)\vartheta(|p|-\Lambda_0)\vartheta(\Lambda-|p|)$,
their low-energy part $\Psi_l(p)\equiv\Psi(p)\vartheta(\Lambda_0-|p|)$
and integrate out the high-energy part of the fermion spectrum,
$\vartheta(t)$ being the usual Heaviside's step distribution.

The above decomposition of the fermion spectrum
should be done covariantly, {\it i.e.}
in terms of the full Euclidean Dirac's normal operator,
\be
\di \equiv i (\not\!\partial + \tau_3 \Phi + \tau_1 H)\ .\la{eudir}
\ee
As we want to investigate  the
$\tau$-symmetry breaking by fermion condensation, in what follows we can
neglect the high-energy components of the auxiliary boson fields,
what is equivalent to adopt the mean-field or large $N$ approximations.
Then the low-energy Euclidean Lagrange density, which accounts for
low-energy fermion and boson fields, can be written as the sum of
the Euclidean counterpart of the classical Lagrange density
\gl{aux} and the induced one-loop contribution, arising by
functional integration of the high-energy fermion field components:
namely,
\be
{\cal L}^{(5)}_{\rm low} (\overline{\Psi}_l,\Psi_l,\Phi, H) =
{\cal L}^{(5)}_E (\overline{\Psi}_l,\Psi_l, \Phi, H) +
\Delta {\cal L}^{(5)} (\Phi, H)\ , \la{lowl}
\ee
where the classical Euclidean Lagrange density reads
\be
{\cal L}^{(5)}_E (\overline{\Psi}_l,\Psi_l,\Phi, H) =
\overline{\Psi}_l\di \Psi_l + \frac{N \Lambda^3}{g_1}\,\Phi^2
+ \frac{N \Lambda^3}{g_2}\,H^2\ .
\la{auxeucl}
\ee
The one-loop contribution of high-energy fermions is given by
\ba
&&\Delta {\cal L}^{(5)} (\Phi, H)=
-(N/2)\,\mbox{tr}\,\langle X|{\rm A}|X\rangle\la{leff}\ , \\
&&{\rm A} \equiv \vartheta(\Lambda^2 - \di^\dagger\!\di)
\ln\frac{\di^\dagger\!\di}{\Lambda^2} - \vartheta(\Lambda_0^2 -
\di^\dagger\!\di) \ln\frac{\di^\dagger\!\di}{\Lambda_0^2}\ ,
\nonumber 
\ea 
where the symbol [ tr ] stands for the trace over
spinor and internal degrees of freedom. In the latter operator [
${\rm A}$ ] we have incorporated the cut-offs which select out the
above defined high-energy region  \ci{AnBo}. For $n=5$ we eventually
found \cite{aags} 
\ba &&\Delta {\cal L}^{(5)} (\Phi,
H)\stackrel{\Lambda\rightarrow\infty}{\sim}\no
&&\frac{N\Lambda}{4\pi^3}\left[\,(\partial_A \Phi)^2 + (\partial_A
H)^2\,\right]-\frac{N\Lambda^3}{9\pi^3}\left(\Phi^2 + H^2\right)
+\frac{N\Lambda}{4\pi^3}\left(\Phi^2 + H^2\right)^2\ .\la{a12} 
\ea
\medskip
Now, let us switch on gravity as described by the metric field
$g_{AB} (X)$ on a five dimensional Riemannian manifold ${\cal M}_5$
which is called the base space. The capital Latin indexes $A,B,C,\ldots $
are usually called the base indexes or holonomic indexes.
The Action is appropriately normalized
with the help of the determinant of this metric,
\be
{S}(\Phi, H, \overline{\Psi}_l, \Psi_l,g) = \int_{{\cal M}_5} d^5 X
\sqrt{g} \ \left[ {\cal L}^{(5)}_{\rm fermion} +
{\cal L}^{(5)}_{\rm boson}\right]\ ;
\quad
g \equiv \mbox{\rm det}(g_{AB})\ , \label{invact}
\ee
where the fermion (spinor) and boson (scalar and gravity) parts of the Lagrange density
will be  defined here below.
Namely, for the Lagrange density \gl{auxeucl} the form invariant under diffeomorphisms
of the part bilinear in fermion fields is given in terms of the pentad--fields
or {\it f\"unfbeine} $e^i_A(X)$,
which connect locally the curved manifold ${\cal M}_5$ with the flat
space with  Euclidean signature, so that
\be
e^i_A(X) e^i_B(X)  = g_{AB}(X)\ ;\quad
e_i^A(X) e_i^{B}(X)  = {g}^{AB}(X)\ ;\quad
e^i_A(X) e_j^A(X) = \delta^i_j\ ,
\ee
where the  Euclidean frame Latin small indexes, also called anholonomic,
run from one to five, in such a way that
\be
\{\widehat\gamma_j, \widehat\gamma_k\} =
2\delta_{jk}\,\left({\bf 1}_4\otimes{\bf 1}_2\right)\ .
\ee
The invariant spinor Lagrange density then reads
\ba
{\cal L}^{(5)}_{\rm fermion} (\overline{\Psi}_l,\Psi_l,\Phi, H, g) &=&
i\overline{\Psi}_l\left[\,\widehat\gamma_k e^A_k \left(\partial_A + \omega_A\right)
+ \tau_3 \Phi +\tau_1 H\,\right] \Psi_l \no
&=&
i\overline{\Psi}_l\left(\,\!\not\!\nabla
+ \tau_3 \Phi + \tau_1 H\right) \Psi_l\no
&\equiv&
\overline{\Psi}_l\ddi\Psi_l\ , \la{aux1}
\ea
where the spin connection $\omega_A$ can be represented in terms of the
{\it f\"unfbeine} and the affine connection
\be
\Gamma^{\,C}_{AB} = \frac12
g^{CD}\left(\partial_A g_{BD} + \partial_B g_{AD}
- \partial_D g_{AB}\right)\ .
\ee
The spin connection has the following form
\begin{equation}
\omega_A \equiv \frac18\,[\,\widehat\gamma_i, \widehat\gamma_j\,]\,\Gamma^{ij}_A\ ;\quad
\Gamma_A^{ij}=e^{Bj} \Gamma_{AB}^{\,C} e_C^i -
e^{Bj} \partial_A e_B^i\ .\label{spincon}
\end{equation}
As before in eq.~\gl{auxeucl}, the fermion self-interaction is
induced by those Yukawa--like vertexes which are
linear in the auxiliary scalar fields $\Phi, H$. We remind that the
symmetry breaking phase arises when the classical scalar interaction
compensates large contributions $\sim \Lambda^3$ in the low-energy effective
action \gl{a12} induced by high-energy fermions.
When the bare dynamical gravity is added, one is expected
to find similar large contributions $\sim \Lambda^5$ and
$\sim \Lambda^3$
in the one-loop effective Action, which may be tuned to end up with a
gravitational dynamics not
severely suppressed by a very high cosmological constant and a
very small Newton's constant. As we will see the zeroth- and first-order
Seeley--Gilkey
coefficients $a_0, a_2$ contain respectively the invariant measure
$\sqrt{g}$ and the scalar curvature $R\,,$
where the curvature scalar is defined in the
conventional way \cite{pauli,land}. Actually, let us denote with
\ba
&&{R^A}_{BCD} =
\partial_C \Gamma_{BD}^A -
\partial_D \Gamma_{BC}^A +
\Gamma_{BD}^E \Gamma_{EC}^A -
\Gamma_{BC}^E \Gamma_{ED}^A\ ;
\label{Riemann}\\
&&R_{BD} \equiv {R^A}_{BAD}\ ;\quad
R \equiv g^{BD}R_{BD}
\ea
the Riemann curvature tensor, the Ricci tensor and the scalar curvature
respectively.

Now, if we aim to compensate all the one-loop large
contributions as induced by the fermionic matter, 
we must add the classical bosonic Euclidean Lagrange
density
\be
{\cal L}^{(5)}_{\rm boson} (\Phi, H, g) = N \Lambda^3
\left(\frac{\Phi^2}{g_1} + \frac{H^2}{g_2}\right) - {\Lambda\over
{\cal G}}\left( \varepsilon \frac{R}{2} - \lambda_0\right)\ , \label{bare}
\ee
We notice that in eq.~(\ref{bare}) the quantity $\lambda_0 \sim \Lambda^2$ 
stands for the bare cosmological constant of the five dimensional
universe. In turn, the Newton--like constant ${\cal G} \sim
1/\Lambda^2$ does specify the strength of the five dimensional
gravitational interaction and its order of magnitude will be
suitably tuned later on. 

Yet, with the help of the factor \mbox{$\varepsilon
= 0, \pm 1$} we reserve ourselves the possibility 
to regulate different physical options for bare 
gravity: namely, for \mbox{$\varepsilon = \pm 1$} the bare Einstein-Hilbert
Action is admittedly generated from a more fundamental theory. 
On the one hand, it turns out that either this bare Action
screens the induced gravity contribution for \mbox{$\varepsilon = -1\,,$} 
just like in the spirit of phenomenological supersymmetry, 
or, conversely, it enhances the induced gravity effect, for \mbox{$\varepsilon=1\,,$} 
so that the five dimensional gravity appears to be very weak. 
Both cases do imply some
particular dynamics beyond the compositeness scale responsible for the 
formation of a pure gravitational interaction. 
On the other hand, the choice of \mbox{$\varepsilon = 0$} does give rise 
to another scenario, in which gravity is entirely induced by fermionic matter, 
{\it i.e.} by proto-fermions in the present model. 
In this latter case, we see how the five dimensional gravitational fields will 
certainly result to be very weak as for \mbox{$\varepsilon =1\,.$} 
We find this option quite interesting and we shall estimate the related scales 
of the five dimensional bulk physics, providing eventually
the usual Newton's gravity in our brane universe.

Now we want to calculate the low-energy effective Action 
in the curved five dimensional space owing to the presence of gravity. 
To this purpose, let us
define the elliptic second-order operator having the
same spectrum as the original Dirac operator in \gl{aux1}.
We use the conjugation property
$\ddi^\dagger = \tau_2\ddi\tau_2$ which relates the diffeomorphism
and frame covariant Dirac operator $\ddi$ to its hermitian
conjugate with respect to the invariant scalar product, as defined
by the invariant measure of eq.\gl{invact}.
Thus, the diffeomorphism and frame
covariant Euclidean Dirac operator
is still a normal operator, which has to be implemented in order to get
a real effective Action
and to define the spectral cut-offs with the help of the positive operator
\ba
\ddi^\dagger\ddi &=& (i\na)^2 +
\Phi^2(X)  + H^2(X)   - \tau_3 \!\not\!\partial \Phi(X)
 - \tau_1 \!\not\!\partial H(X)\ . \la{quad}
\ea
This elliptic second-order differential positive operator can be re-expressed in a quite
general form \ci{cogia},
\be
\ddi^\dagger\ddi\ =\ -\ g^{AB}(X)D_A D_B + {\cal M}^2(X)\
\equiv\ -\ D^2 + {\cal M}^2(X)
\label{eq:KG-heat}
\ee
where ${\cal M}^2(X)$ is a matrix--valued multiplicative ({\it i.e.} non--differential)
operator, whereas
the full covariant derivative for a mixed quantity $f$ having anholonomic
(left understood) and holonomic indexes is given by
$D_A f^C = (\partial_A + \omega_A)f^C + \Gamma^C_{AB} f^B\,.$
Notice that the full covariant Laplace operator $D^2$
reduces to the Laplace-Beltrami operator when acting on
a scalar quantity
\be
D^2f(X)=g^{AB}(X)\partial_A\partial_B f(X)+
\frac{1}{\sqrt{g(X)}}\,\partial_A
\left[\,g^{AB}(X)\sqrt{g(X)}\,\right]\partial_B f(X)\ .
\ee
Explicit evaluation yields
\be
{\cal M}^2(X) =
\frac{R}{4} + \Phi^2(X)  + H^2(X) -
\tau_3 \!\not\!\partial \Phi(X) -
\tau_1 \!\not\!\partial H(X)\ .
\ee
where the identity matrix $\widehat{\bf 1}\equiv {\bf 1}_4\otimes{\bf 1}_2$
is always left understood.
Now, let us calculate the effective low energy Lagrange density
induced by high energy proto--fermions.
One can see that, in fact, the scale anomaly only contributes
into  $\Delta {\cal L}^{(5)}$, {\it i.e.} that part which depends upon the scales.
Thus, equivalently,
\be
 \Delta {\cal L}^{(5)} (\Phi, H, g) = N \int^{\Lambda}_{\Lambda_0}
\frac{dQ}{Q}\
\mbox{tr}\langle X|
\vartheta(Q^2 - \ddi^\dagger\ddi)|X\rangle\ .
\la{scan}
\ee
As we assume that the scalar fields carry momenta much smaller than the lower scale
$\Lambda_0$, then the diagonal matrix element
in the RHS of eq.\,\gl{scan} can be
calculated with the help of the derivative expansion of the
representation \ci{AnBo}
\be
\mbox{tr}\langle X|\vartheta(Q^2 - \ddi^\dagger\ddi)|X\rangle =
\int_{-\infty}^{+\infty}\frac{d t}{2\pi i}\
\frac{\exp\{i t\}}{t - i\varepsilon}\ \mbox{tr}\langle X|
\ \exp\left\{- i\,\ddi^\dagger\ddi/{Q^2}\right\}|X\rangle
\label{intrep}
\ee
where $X$ belongs to the base space,
which is supposed to be a Riemannian manifold of dimension $n\,.$
The trace in eq.~(\ref{intrep}) can be calculated by means 
of the heat kernel method (see its review \ci{vasil}), as shown in Appendix A.
For $n=5 $, only three heat kernel coefficients  at most
are proportional to non--negative powers of the large parameter $Q$,
\ba
&&\mbox{tr}\langle X|\vartheta(Q^2 - \ddi^\dagger\ddi)|X\rangle \approx \frac{Q^5}{15\pi^3}
- \frac{Q^3}{3\pi^3}\,\left[\,\Phi^2(X) + H^2(X) + \frac{R(X)}{12}\,\right] \no
&& +\ \frac{Q}{4\pi^3}\left\{
\partial_A\Phi(X)\partial^A\Phi(X) +
\partial_A H(X)\partial^A H(X) +
[\,\Phi^2(X)+H^2(X)\,]^2\right.\no
&&\left. +\ \frac{1}{6}\,R(X)[\,\Phi^2(X)+H^2(X)\,]
- \frac13\,D^2[\,\Phi^2(X)+H^2(X)\,] - \frac{1}{60}\,D^2 R(X)
\right.\no
&&\left. -\ \frac{7}{720}\,
R_{ABCD}(X)R^{ABCD}(X) - \frac{1}{90}\,
R_{AB}(X)R^{AB}(X) + \frac{R^2(X)}{144}\right\}\ ,
\la{elem}
\ea
where, for large scales $\Lambda_0\ll Q < \Lambda$, the neglected terms
rapidly vanish.
Inserting the RHS of eq.\,\gl{elem} in eq.\,\gl{scan} and taking into account
that $\Lambda_0 \ll \Lambda$, one can neglect the $\Lambda_0$--dependence
and find, up to a total penta--divergence,
\ba
\Delta {\cal L}^{(5)} (\Phi, H, g)\stackrel{\Lambda\rightarrow\infty}{\sim}&&
\frac{N\Lambda^5}{75\pi^3}-\frac{N \Lambda^3}{9\pi^3}
\left(\Phi^2 + H^2 + \frac{R}{12}\right)\no
+&&\frac{N \Lambda}{4\pi^3}
\left\lbrace\,\partial_A \Phi\,\partial^A \Phi +
\partial_A H \partial^A H\right.\no
+&&\left.\left(\Phi^2 + H^2\right)^2 +
 \frac{R}{6}\,\left(\Phi^2 + H^2\right) +
\frac{ R^2}{144}\right.\no
-&&\left.\frac{1}{90}\,
R_{AB}R^{AB}- \frac{7}{720}\,
R_{ABCD}R^{ABCD}\,\right\rbrace\ .
\la{a123}
\ea
Although the actual values of the coefficients might
be regulator--dependent, as already noticed,
the coefficients of the kinetic
and quartic terms of the effective low energy Lagrange density are definitely equal,
no matter how the latter is
obtained from the basic Dirac operator of eq.\,\gl{eudir}.


\section{Scalars and gravity: classical configurations}

The interplay between different operators in the low-energy
Lagrange density \gl{lowl} may lead to different dynamical
regimes, depending of $\tau$-symmetry breaking. 
The low-energy Euclidean Lagrange density can be cast in the form
\ba
{\cal L}_{\rm low}^{(5)} (\overline{\Psi}_l,\Psi_l,\Phi, H, g) &\equiv&
{\cal L}^{(5)}_{\rm fermion} (\overline{\Psi}_l,\Psi_l,\Phi, H, g) +
{\cal L}^{(5)}_{\rm boson} (\Phi, H, g) +
\Delta {\cal L}^{(5)} (\Phi, H, g)\no
&=& i\overline{\Psi}_l(X)\left[\,\na
+ \tau_3 \Phi(X) + \tau_1 H(X)\,\right]\Psi_l(X)\no
&+& \frac{N\Lambda}{4\pi^3}\,\left\{\partial_A \Phi(X)\partial^A \Phi(X)+
\partial_A H(X)\partial^A H(X)\right.\no
&-&\left. 2 \Delta_1\Phi^2(X)
- 2\Delta_2 H^2(X)\right\} - \frac{\Lambda}{2 \kappa \cal G}\left\lbrace\,
R(X) - 2\lambda\,\right\rbrace\no
&+& \frac{N\Lambda}{4\pi^3}\,\left[\,\Phi^2(X) + H^2(X)\,\right]\left\{
\Phi^2(X) + H^2(X) + {R(X)\over 6}\right\}\no
&+&\frac{N\Lambda}{2880\pi^3}\left\lbrace 5R^2(X) -
8R_{AB}(X)R^{AB}(X)\right.\no
&-& \left.
7R_{ABCD}(X)R^{ABCD}(X)\right\rbrace
\la{lowmin}
\ea
where the two mass scales $\Delta_i$ characterize the
deviations from the critical point
\be
\Delta_i(g_i)
=\frac{2\Lambda^2}{9g_i}\left(g_i-g_i^{\rm cr}\right);\quad g_1^{\rm cr} = g_2^{\rm cr}
= 9\pi^3 \ .
\ee
The {\sl dressed} cosmological  constant $\lambda$ and the
gravitational low-energy dimensionless parameter $\kappa$ 
do arise as a net effect of the interplay between the classical 
and fermion induced contributions\footnote{We remark that the
combination of terms quadratic in the scalar and tensor curvatures
does not form the Gauss-Bonnet gravity interaction \ci{corra} and, as
well, the scalar curvature ``mass'' $R/6$ for scalar matter fields does
not imply an additional 
conformal symmetry of Klein-Fock-Gordon equations for the 
five-dimensional manifold \ci{tagir} .}. 
If we introduce some average compositeness scale
\be
\Lambda^3_c \equiv N\Lambda^3/54\pi^3, \la{compo}
\ee
together with a five dimensional counterpart of the Planck scale
\be
{\hbar c\Lambda}/{{\cal G}\kappa} \equiv M^3_*\ ,
\la{plfive}
\ee
then we can easily obtain 
\be
\lambda=\kappa \lambda_0+\frac{N\Lambda^4}{75\pi^3}\,{\cal G} \kappa \ ,
\ee
together with the relationship
\be
 {\kappa}
= \frac{1}{\varepsilon + N\Lambda^2{\cal G}/54\pi^3}\,\quad 
\varepsilon\kappa=1-\left({\Lambda_c\over M_*}\right)^3
\label{kappa}
\ee
%
%
where $\kappa$ must be positive for the gravitational
interaction to be attractive.
We shall be interested in different scenarios which one can 
qualitatively specify as follows:\\ 
\begin{itemize}
\item[(A)]
{\tt Fundamental gravity},\\
when the bare and dressed gravitational couplings are comparable
\be
\varepsilon  = 1\ ,\qquad 
\kappa \simeq 1\ ,\qquad
\mbox{\rm so that}\qquad  
\Lambda_c \ll  M_*\ .
\label{fundgrav}
\ee
\item[(B)]
{\tt Induced gravity},\\ 
when the bare gravitational Action is
either absent  or irrelevant whilst the fermion induced
gravitational Action is dominant
\be 
\varepsilon  = 0\ ,\qquad\mbox{and/or}\qquad  
\kappa \ll 1\ ,\qquad 
\mbox{\rm so that}\qquad 
\Lambda_c \simeq  M_*\ . 
\label{indgrav}
\ee 
\item[(C)]
{\tt Strong gravity},\\
when the bare gravitational Action is dominating
with respect to the fermion induced gravitational Action
\be
\varepsilon  = -1\ ,\qquad 
\kappa \ma 1\ ,\qquad
\mbox{\rm so that}\qquad  
\Lambda_c >  M_*\ . 
 \label{strongrav}
\ee 
\end{itemize}
Eventually, we shall demand for all scales to be much less than 
the five dimensional Planck-like mass, otherwise 
our effective model would go beyond its own essence and
must be replaced by the still unknown genuine theory of quantum gravity. 
This criterion rather rules out the scenario $(C)$ with $\kappa \gg 1,\quad
\Lambda_c \gg  M_*$ as the very concept of composite
particles and fields far above the Planck scale certainly needs to engage 
quantum gravity. This is why we will not focus on it keeping the side
of a weak quasi--classical gravity.

In what follows we assume that all the matter effective couplings substantially reduce 
the cut-off scale $\Lambda$ to a low
energy scale $M \ll \Lambda$ and, to be definite, we take $\Delta_1=M^2\,.$
Thus, in
order to make the different parts of the equations of motion comparable in mass scales,
we shall suitably tune in the sequel the different
couplings of our five dimensional model.
In particular, the bare cosmological constant must be negative 
$\lambda_0 <0$ in order to provide
a small constant $\lambda \ll \Lambda^2\,.$

Let us obtain brane-like solutions which describe the localization
along the fifth coordinate $z$. In this paper we are interested in the
static effects of particle trapping on a brane and this is why we shall
consider here only the flat four dimensional space-time, with
Euclidean signature to simplify our calculations.
As a consequence, we restrict ourselves to analyze
the quasi-flat Riemannian metric, the invariant line element
of which can be suitably chosen as follows
\ba
ds^2 = g_{AB}(X)\,dX^A\,dX^B
= \exp\{-2\rho(z)\}\,dx_\mu dx_\mu + dz^2
 \label{metr}
\ea
with the Euclidean signature. The physical motivations for the above choice will
be clear later on.
The related invariant volume factor is $\sqrt{g} = \exp\{-4\rho(z)\}$.
The scalar and gravity parts of the low-energy
Lagrange density \gl{lowmin} for this metric reduce to 
(see the expressions for curvature terms in Appendix B)
\ba
&&{\cal L}^{(5)}_{\rm boson} (\Phi, H, g) + \Delta {\cal L}^{(5)} (\Phi, H, g)\ =\no
&& ({N\Lambda}/{4\pi^3})\,\exp\{2\rho(z)\}\,
\left[\,\partial_\mu \Phi(X)\partial_\mu \Phi(X) + \partial_\mu H(X) \partial_\mu H(X)\right]
+ ({N\Lambda}/{4\pi^3})\ \times\no
&& \times \left\lbrace
 \left[\,\partial_z \Phi(X)\,\right]^2 +\left[\,\partial_z H(X)\,\right]^2
- 2\Delta_1\Phi^2(X) - 2\Delta_2 H^2(X)
+ \left[\,\Phi^2(X) + H^2(X)\,\right]^2\right\rbrace\no
&& +\ ({N\Lambda}/{3\pi^3})\,\left[\,\Phi^2(X) + H^2(X)\,\right]
\left\{\rho''(z) - (5/2)\,[\,\rho'(z)\,]^2\,\right\}\no
&& +\ ({\Lambda/\kappa{\cal G}})\,\left\{- 4\rho''(z)
+\ 10\,[\,\rho'(z)\,]^2\, +\ \lambda\right\}\no
&& +\ ({N\Lambda}/{120\pi^3}) \left\lbrace
2[\,\rho''(z)\,]^2  - 36 \rho''(z)[\,\rho'(z)\,]^2 + {45}[\,\rho'(z)\,]^4\right\rbrace
\la{confm}
\ea
where $\rho^\prime(z)\equiv d\rho/dz\,,\  \rho^{\prime\prime}(z)\equiv d^{\,2}\rho/dz^2\,.$
Let us search for classical solutions in the form of a metric \gl{metr} in the
gravitational weak coupling regime in which we assume that all along
the large extra dimension we have
\be
|\rho^\prime (z)|/M = \O(\kappa)\ ,\qquad
|\rho^{\prime\prime} (z)|/M^2 = \O(\kappa)\ .
\ee
This means that the conformal function $\rho(z)$ is
slowly varying over a distance, in the  extra dimension, 
of the order of magnitude of the typical Compton wavelength of the low energy particles.
This requirement, together with $0\mi \kappa\ll 1$ and $M\ll \Lambda$,
yields that the last
line in eq.~\gl{confm}, which corresponds to the terms
quadratic in the curvature scalar and curvature tensors
of the Lagrange density \gl{lowmin},
will be sub--leading and negligible with respect to the other terms.
Nevertheless, just for completeness, we shall display the contribution 
of those terms into the equations of motion in Appendix C. 

Within the present gravitational
weak coupling approximation,
the leading dynamics is mainly determined by the
Einstein--Hilbert Action, on the five dimensional Riemannian
manifold ${\cal M}_5$,
and its coupling to the scalar matter fields.
As shown in Appendix D, the field equations become
\be
R_{AB} -\frac12\,g_{AB}\left(R - 2\lambda\right)\
\equiv G_{AB}\ +\ \lambda\,g_{AB}\ =\
\frac{N\kappa{\cal G}}{2\pi^3}\,t_{AB}
\ee
where the normalized energy--momentum tensor of the scalar matter reads
\ba
t_{AB}\ &\equiv&\ ({4\pi^3}/{N\Lambda})\,T_{AB}\ \equiv\
\partial_A \Phi\,\partial_B\Phi +
\partial_A H\,\partial_B H\no
&-&\ \frac12\,g_{AB}\left[\,
\partial_C \Phi\, \partial^C \Phi +
\partial_C H\,\partial^C H - 2 \Delta_1\Phi^2
- 2\Delta_2 H^2 + \left(\Phi^2 + H^2\right)^2\,\right]\no
&+&\  \frac16\left(R_{AB} - \frac12\,g_{AB}\,R
+ g_{AB}\,D^C\partial_C  - D_B\,\partial_A\right)
\left(\Phi^2 + H^2\right)\ .
\ea
The equations of motion for the scalar fields read
\ba
&&2\left[\Delta_1  - \Phi^2 - H^2\,\right]\Phi\ =\
\left(\frac{R}{6}\ -\ \frac{1}{\sqrt{g}}\,
{\partial}_C \sqrt{g} \ {g}^{CD} {\partial}_D\right) \Phi\ ,\no
&&2 \left[\,\Delta_2  - H^2 - \Phi^2\,\right]H\ =\
\left(\frac{R}{6}\ -\ \frac{1}{\sqrt{g}}\,
{\partial}_C \sqrt{g} \ {g}^{CD} {\partial}_D\right) H\ .
\la{phivar1}
\ea
In the case of a quasi--flat metric \gl{metr} and for
kink--like profiles of the vacuum expectation values of the scalar fields
\mbox{$\langle\Phi(X)\rangle_0 = \Phi(z),$} \mbox{$\langle
  H(X)\rangle_0 = H(z)$}
one finds the following equations: namely,
\ba
G_{\alpha\alpha}(z)\ +\ g_{\alpha\alpha}(z)\,\lambda\ &=&\
\left({N\kappa{\cal G}}/{2\pi^3}\right)\,t_{\alpha\alpha}(z)\ ,\\
G_{55}(z)\ +\ \lambda\ &=&\
\left({N\kappa{\cal G}}/{2\pi^3}\right)\,t_{55}(z)\ ;
\ea
which respectively reduce to
\ba
\rho^{\,\prime\prime} - 2\rho^{\,\prime\,2} - \frac13\,{\lambda} &=&
\frac{N\kappa{\cal G}}{12\pi^3}\left\lbrace
\Phi^{\,\prime\,2} + H^{\,\prime\,2} - 2 \Delta_1\Phi^2 - 2\Delta_2 H^2
+ \left(\Phi^2 + H^2\right)^2\right.\no
&+& \left.\left( \rho^{\,\prime\prime} - 2\rho^{\,\prime\,2}
- \frac13\,\frac{d^{\,2}}{dz^2} + \rho^{\,\prime}\,\frac{d}{dz}\right)
\left(\Phi^2 + H^2\right)\right\rbrace
\label{eqgrav1}
\ea
\ba
2\rho^{\,\prime\,2} + \frac13\,{\lambda}  &=&
\frac{N\kappa{\cal G}}{12\pi^3}\left\lbrace
\Phi^{\,\prime\,2} + H^{\,\prime\,2} + 2 \Delta_1\Phi^2 + 2\Delta_2 H^2
\right.\no
&-& \left.\left(\Phi^2 + H^2\right)^2 + \left(2\rho^{\,\prime\,2}
- \frac43\,\rho^{\,\prime}\,\frac{d}{dz}\right)
\left(\Phi^2 + H^2\right)\right\rbrace\ .
\label{eqgrav2}
\ea
The field equations for the scalar matter become
\be
\Phi^{\,\prime\prime} =
2 \Phi\left(\Phi^2 + H^2\right) - 2\Delta_1\Phi
+ 4\rho^\prime\, \Phi^\prime
+ \frac{2}{3}\,\Phi\left(2\rho^{\,\prime\prime}
- 5\rho^{\,\prime\,2}\right)\ ,\label{eqscalar1}
\ee
\be
H^{\,\prime\prime} =
2 H\left(\Phi^2 + H^2\right) - 2\Delta_2 H
+ 4\rho^\prime\, H^\prime
+ \frac{2}{3}\,H\left(2\rho^{\,\prime\prime}
- 5\rho^{\,\prime\,2}\right)\ .
\label{eqscalar2}
\ee

Now, it turns out to be convenient to put forward the role
of the low--energy relevant mass scale $\Delta_1=M^2$. This can be better
achieved after a suitable redefinition of the gravitational low--energy
dimensionless parameter
\be
{\overline \kappa}\ \equiv\ \frac{N\kappa}{6\pi^3}\,M^2{\cal G}\ \ll\ 1
\label{newcoup}
\ee
and of the cosmological constant
\be
\lambda\ \equiv\ 3{\overline \kappa}\,\lambda_{\rm
  eff}\ .
\label{coseff}
\ee
It is worthwhile to notice that the sum of Eqs.~\gl{eqscalar1} and \gl{eqscalar2}
does not include the five dimensional rescaled cosmological constant $\lambda_{\rm eff}$
and reads
\be
\rho^{\prime\prime}  = \frac{{\overline \kappa}}{M^2}
\left\{\Phi^{\,\prime\,2} + H^{\,\prime\,2} +
\frac12\left(\rho^{\prime\prime}
- \frac13\,\frac{d^{\,2}}{dz^2} - \frac13\,\rho^{\,\prime}\frac{d}{dz}\right)
\left(\Phi^2 + H^2\right)\right\}\ .
\label{eqgrav05}
\ee
On the other hand, eq.~\gl{eqgrav2} can be rewritten in the form
\ba
2M^2\lambda_{\rm eff}\  &=&
\Phi^{\prime\,2}
+ H^{\prime\,2}  + 2 \Delta_1\Phi^2 + 2\Delta_2 H^2
- \left(\Phi^2 + H^2\right)^2\no
&+& \left(2\rho^{\,\prime\,2} - \frac43\,\rho^\prime\frac{d}{dz}\right)
\left(\Phi^2 + H^2\right) - \frac{4M^2}{{\overline \kappa}}\,\rho^{\,\prime\,2}
\label{intcosm}
\ea
which actually represents the integral of motion with
the rescaled five dimensional cosmological constant playing 
the role of an integration constant.
This can be checked by differentiating the above  equation
and taking into account Eqs.~\gl{eqscalar1}, \gl{eqscalar2}
and \gl{eqgrav05}.

We can approximate our field equations
keeping in mind that we still assume
\be
\frac{|\rho^\prime(z)|}{M}\ =\ \O({\overline \kappa})\ =\
\frac{|\rho^{\prime\prime}(z)|}{M^2}\ ,
\label{approx}
\ee
as previously specified. We emphasize that the validity of
perturbation theory in $\overline \kappa$ is endorsed by the choice
of a reference frame where the metric takes the form \gl{metr}. 
For different choices such as, for instance, the conformal metric 
\be
ds^2 = \exp\{-2\sigma(y)\}\,\Bigl(dx_\mu dx_\mu + dy^2\Bigr)\ ;\quad
 \exp\{-\sigma(y)\}dy \equiv dz\ ,
\ee 
the expansion in  $\overline \kappa$
is not uniform and fails for large $z\,.$

On the other hand, the expansion in powers of ${\overline \kappa}$ 
of the solutions of Eqs.~\gl{eqscalar1}, \gl{eqscalar2},
\gl{eqgrav05} and \gl{intcosm} is well defined on the whole
extra dimension $-\infty < z < \infty\,.$
Let us find solutions by expanding in
${\overline \kappa}\ll 1$.
According to our approximation \gl{approx},
taking into account that $\Phi/M = {\O}(1) = H/M\,,$ we obtain in
the leading order
\be
\frac{\rho^{\,\prime\prime}}{M^2}  = \frac{{\overline \kappa}}{M^4}
\left\{\Phi^{\prime\,2} + H^{\prime\,2}
- \frac16\,\frac{d^{\,2}}{dz^2}
\left(\Phi^2 + H^2\right)\right\} + \O({\overline \kappa}^2)\ ,
\label{grav0}
\ee
whereas the cosmological constant to the lowest order reads
\be
\frac{\lambda_{\rm eff}}{M^2}\  =
\frac{1}{2M^4}\left\lbrace\Phi^{\prime\,2}
+ H^{\prime\,2}  + 2 \Delta_1\Phi^2 + 2\Delta_2 H^2
- \left(\Phi^2 + H^2\right)^2\right\rbrace + \O({\overline \kappa})\ .
\label{cosm0}
\ee
The latter equation firmly determines the five dimensional
cosmological constant
$\lambda_{\rm eff}$ in terms of the parameters for the kink--like
solutions in the flat space which read
\ba
&&\Phi^{\,\prime\prime} + 2\,\Phi\left(\Delta_1
- \Phi^2 - H^2\right) = \O({\overline \kappa})\ ,\no
&&H^{\,\prime\prime} + 2H\left(\Delta_2
- \Phi^2 - H^2\right) = \O({\overline \kappa})\ .
\label{scal0}
\ea
To sum up, we have four equations for three functions $\rho(z), \Phi(z), H(z)$
and one integration constant $\lambda_{\rm eff}\,.$
As in \cite{aags} one can
discover two types of kink--like  solutions for
Eqs.~\gl{scal0}: namely,
\ba
&&(J)\quad \Phi_J \equiv \langle\Phi(X)\rangle_0 = M
\mbox{\rm tanh}(M z)\ , \quad  H_J \equiv\langle H(X)\rangle_0
= 0\ ;\la{Jvacuum}\\
&&(K)\quad \Phi_K \equiv \langle\Phi(X)\rangle_0=  M
\mbox{\rm tanh}(\beta z)\ ,\quad H_K \equiv
\langle H(X)\rangle_0 =\mu\,\mbox{\rm sech}(\beta z)\ ,
\label{Kvacuum}
\ea
where
\be
\mu =
\sqrt{2 \Delta_2 - M^2}\ ,\qquad
\beta = \sqrt{M^2 - \mu^2}\ .
\label{mu}
\ee
The solution $(K)$ exists only for $\Delta_2 < M^2 < 2\Delta_2$
and it coincides with the extremum $(J)$ in the limit
$\Delta_2\to M^2/2,\ \mu\to 0,\ \beta\to M$.
Both of them give consistently
\be
 \lambda =
\frac{N{\cal G}\kappa M^4}{4\pi^3} = \frac32 \bar\kappa M^2\ .
\label{cosmcon}
\ee
The last term in \gl{grav0} is originated from
the scalar curvature mass--like term $R/6$
in the Lagrange density
\gl{lowmin} for the scalar fields. To unravel its role, we split
the conformal factor in the following way: namely,
\ba
\rho(z) &=&\rho_1(z) + \rho_2(z)
= \frac{2{\overline \kappa}}{3}
\left(1 +\frac{\mu^2}{2M^2}\right)\ln\mbox{\rm cosh}(\beta z)
+ B z\ ;\label{rhoone}\\
\rho_1^{\,\prime\prime}(z)  &=&
 \frac{{\overline \kappa}}{M^2} \left\lbrace[\,\Phi^\prime(z)\,]^2
+ [\,H^\prime(z)\,]^2\right\rbrace\ ;\quad
\rho_2^{\,\prime\prime}(z)  =
\frac{- {\overline \kappa}}{6M^2}\,{d^2\over dz^2}\,
\left[\,\Phi^2(z) + H^2(z)\,\right]\ ;\no
\rho_1(z) &=& \frac{2{\overline \kappa}}{3}
\left(1 +\frac{\mu^2}{2M^2}\right)\ln\mbox{\rm cosh}(\beta z)
+ \frac{{\overline \kappa}}{6}\left(1 -\frac{\mu^2}{M^2}\right)
\mbox{\rm tanh}^2(\beta z) + B z\ ; \no
\rho_2(z) &=& \frac{-{\overline \kappa}}{6M^2}
\left[\,\Phi^2(z) + H^2(z) - \mu^2\,\right]\
=\ -\ \frac{{\overline \kappa}}{6}
\left(1 -\frac{\mu^2}{M^2}\right)
\mbox{\rm tanh}^2(\beta z)\ .
\ea
This solution is normalized so that for the vanishing integration constant
\mbox{$B \rightarrow 0$} the function $\rho(z)$ becomes even  and
$\rho (0) = 0$. The latter corresponds to the proper normalization
of the {\it 3+1}
metric on the brane $z = 0\,.$

One can see that the scalar curvature term substantially simplifies the
metric factor $\rho(z)$ around the brane. Evidently, this solution approaches,
when $B=0$, the symmetric Anti-de-Sitter (AdS) metric for large $z$: namely,
\be
\rho(z)\stackrel{|z| \rightarrow \infty}{\sim}
 k|z|\ ;\quad k\equiv \frac23\,{\overline \kappa}\beta
\left(1 +\frac{\mu^2}{2M^2}\right)\ \approx\
\frac23\,{\overline \kappa}M\ .
\label{AdSlim}
\ee


\section{Ultra-low energy physics in the matter sector}

Let us summarize the structure \cite{aags} of the spectrum and 
of the interaction of the light states trapped on a brane,  
in the absence of gravity. The kinetic operators 
(second variation of the Action) of the two scalars 
$\Phi(X)$ and $H(X)$ and of the spinor field $\Psi(X)$ 
do exhibit normalizable zero-modes in the extra dimension, 
in the vicinity of the vacuum background \gl{Jvacuum} or \gl{Kvacuum},
at the scaling point $M^2 = \Delta_1 = 2 \Delta_2$ or $\mu = 0\,.$
Those zero-modes $\phi_0(z)$, $h_0(z)$ and $\psi_0(z)$, respectively, 
are localized at the origin of the $z$-axis, with a localization 
width $\sim 1/M$ and, at ultra-low energies, the fluctuations of the
matter fields can be parametrized as follows:  namely,
\ba 
&&\Phi(X) \simeq \langle\Phi(X)\rangle_0 + \phi(x)
\phi_0(z)\ ;\no
&& H(X) \simeq\langle H(X)\rangle_0 +  h(x) h_0(z)\ ;\no
&& \Psi(X) \simeq
\psi(x)\,\psi_0 (z)\ .
\ea 
For these states the ultra-low-energy effective Lagrange density 
(still in the Euclidean space) is generated
\ba
{\cal L}^{(4)} &=&  
i\overline{\psi}(x)\left[\,\not\!\partial + g_f\,h(x)\,\right]\psi(x)
+\frac12[\,\partial_\mu\phi(x)\,]^2
+\frac12[\,\partial_\mu h(x)\,]^2\no
&+& \lambda_1\phi^4(x) + \lambda_2\phi^2(x) h^2(x) + \lambda_3 h^4(x)\ ,
\la{lelag}
\ea
with the ultra-low energy effective couplings given by
\be 
g_f = \frac{\pi}{4} \sqrt{\zeta}\ ,\quad \lambda_1 =
\frac{18}{35}\,\zeta\ ,\quad \lambda_2
=\frac{2}{5}\,\zeta\ , \quad \lambda_3
=\frac{1}{3}\,\zeta\ , \qquad \zeta \equiv \frac{M\pi^3}{\Lambda N}\ .
\la{lowcons}
\ee
Once gravity is switched on, it can be shown\footnote{It
has been firstly analyzed in \ci{misse,missch,galoc2}. A detailed analysis more
suitable for our model will be presented elsewhere.}  that the zero-modes
remain localizable -- see below -- and therefore the AdS vacuum solution 
does not play any dominant role concerning the determination of the coupling 
constants in eq.~\gl{lelag}.

The situation becomes more subtle when light massive states are there. 
In the absence of  gravity,
the deviations off the scaling point towards
the $(K)$ vacuum configuration with $\mu \ll M$ do
produce the masses for the Higgs-like particle and the spinor
particles. Furthermore, the scalar self-interaction is induced
in the form
\ba
&&\Delta {\cal L}_\mu^{(4)} = \frac12\,m^2_h\,h^2(x) + i m_f\,
\overline{\psi}(x)\psi(x) + \lambda_4\,h^3(x)\ ;\no
&&m^2_h = \mu \sqrt2\ ;\qquad
 m_f = \frac{\pi}{4}\,\mu\ ;\qquad 
\lambda_4 = \mu \sqrt{\zeta}\ .
\la{mumass}
\ea
On the one hand, we see that all interaction vertexes 
are governed by the parameter $\zeta \sim M/\Lambda$ and
if $\zeta \ll 1$ the scalar matter essentially 
decouples from the fermion sector and 
does not interact without gravity. However, the
parameter $\zeta$ is not properly fixed  if gravity is not present 
and, in general, it is constrained by experimental bounds.
On the other hand, the masses of Higgs-like scalar and fermions are 
controlled by the ultra-low scale $\mu$
independently of $\zeta\,.$ In the spirit of the Top-mode Standard Model
\ci{Topmode} one expects that the heaviest quark mostly
contributes into dynamical symmetry breaking, thereby driving to the scale
$\mu \sim m_{\rm top}\sim 200$ GeV, {\it i.e.} the order of 
magnitude of the electroweak symmetry breaking scale \ci{PDG}.

Once the five dimensional gravity is switched on and the background geometry 
becomes an anti-de Sitter one, it turns out that light massive scalar and fermion 
1-particle states belong, in fact, to the continuous part of the Hamiltonian spectrum,
the latter ones being strongly localized on the brane albeit possessing a tiny 
non-vanishing tail for $|z| \gg 1/k\,.$ 
The origin of this phenomenon can be schematically sketched in terms of
the mass operator for a one dimensional scalar field living on the fifth 
dimension (see eq.~\gl{confm} and \ci{aags}): namely,
\ba
\exp\{-2\rho(z)\} {\bf M}_z &\equiv& -\ \partial_z[\,\exp\{- 4
\rho(z)\}\partial_z\,] + {\bf  V}(z) \exp\{- 4 \rho(z)\}\ ;\no   
{\bf M}_z\,\varphi(z) &=& m^2 \varphi(z)\ ,
\ea
where ${\bf  V}(z)$ is, in general, some matrix-valued potential 
well giving rise to bound states, in the absence of gravity,
and approaching the energy $4M^2$ for large $z\,.$ 
The mass eigenvalue $m$ stands for the mass of a scalar Higgs-like 
1-particle state.
After the suitable redefinition $\varphi(z) = \widetilde\varphi(z)\,
\exp\{2\rho(z)\}\,,$ the eigenvalue problem is transformed into
a zero-mode condition\quad$\widetilde{\bf M}_z\,\widetilde\varphi(z)=0\,,$ 
where the modified mass operator takes the conventional form,
\be
\widetilde{\bf M}_z = -\ \partial_z^2 + \widetilde{\bf  V}(z) 
- m^2 \exp\{2\rho(z)\}\ \equiv\ -\ \partial_z^2 + {\bf  W}(z)\ , 
\la{massop2}
\ee
whereas $\widetilde{\bf  V}(z)\simeq {\bf  V}(z)$ uniformly 
up to the leading order in ${\overline \kappa}\,.$ 
Thus we see how the only important difference in the spectral
problem, in comparison with the five dimensional flat space, 
does consist in the appearance of the conformal factor, 
which is boundless increasingly when $|z| \gg 1/k\,.$ 
The very last term in eq.~\gl{massop2} becomes now a piece of the 
modified potential ${\bf  W}\,,$ whilst the mass eigenvalue $m$ 
just appears as a coupling constant.
 
Evidently, this very last piece, being negative, 
makes the modified potential unbounded from below: instead of a potential
well,  some finite barrier just arises. 
As a consequence, each light massive
state with $0 < m^2 \ll 4M^2$ is at most
quasi-stationary. Nonetheless, let us assign massive scalar particles in the
brane world to be described by normalizable wave packets with a width
of the order $1/M\,.$ The latter ones are not, of course,  
exact eigenfunctions of the mass operator \gl{massop2} but may be
taken as close as possible, {\it e.g.} with the help of a variational principle, 
to the solutions in the vicinity of the brane $|z| \ll 1/k\,.$ 
For instance, one can start with bound state wave functions in the absence of
gravity. Then quantum mechanics predicts the decay rates of such a kind of
bound states, due to quantum tunneling. 
The quasi-classical 
probability of barrier penetration is governed by the change in 
the sub-barrier wave function between the two turning points for classical trajectories,
say, $0 < z_0 < z_1\,.$ These turning points can be estimated to be
\be
z_0 = \frac{C}{M}\ ,\quad C \simeq 1\ ; \qquad z_1 \simeq \frac{1}{k} 
\ln\frac{2M}{m}\ ,
\ee
so that $z_1 \longrightarrow \infty$ for the zero-modes 
$m \longrightarrow0\,.$ Taking these relations into account, one finds the suppression
factor for quantum tunneling
\be
\exp\left\{- \int^{z_1}_{z_0} dz' \sqrt{{\bf  W}(z')}\,\right\} \simeq 
\exp\left\{- z_1 \cdot 2M\right\} \simeq 
\exp\left\{- \frac{3}{{\overline \kappa}}\ln\frac{2M}{m}\right\}\ , 
\la{suppr}
\ee
where the limit ${\bf  W} \longrightarrow 4M^2$ for $z_0 \ll z'\ll
z_1$  and the definition \gl{AdSlim} have been used.

As we shall discuss in the next Section, some reasonable order of
magnitude for the weak gravitational constant is 
${\overline \kappa}\leq 10^{-8}\,,$ 
so that it is clear that such a suppression factor \gl{suppr} is
extremely small and spoils in practice any chance for matter to
disappear from the brane. We also notice that the suppression factor 
is essentially non-perturbative and cannot be recovered as an expansion 
in powers of ${\overline \kappa}$ -- see {\it e.g.} \ci{CGS}  for a
complete discussion of the similar case of the Stark effect. 

Now we can outline our further strategy in 
calculating gravitational effects on the mass spectrum and coupling
constants of light particles living on the brane. First, particle wave
functions at the lowest order are taken from the flat space limit. 
Second, only perturbation theory in powers of ${\overline \kappa}$ is involved 
in corrections of particle characteristics due to non-trivial 
gravitational background. In this way, our approach is able to
implement the systematic perturbative expansion which provides
light particle wave packets localized upon the brane, to any finite
order in the expansion in powers of ${\overline \kappa}\,.$
A more detailed and rigorous treatment is postponed to the forthcoming paper. 
  

\section{Newton's constant and parameters}

One can find the relation between the five dimensional and brane gravity
constants using the factorized Riemannian metric
\be
ds^2 = \exp\{-2\rho(z)\}\,g_{\mu\nu} (x) dx^\mu dx^\nu + dz^2\ .
\label{metr2}
\ee
For this metric, according to eq.~\gl{scalarcur5} of Appendix B,
the effective four dimensional Einstein-Hilbert Action,
at the leading order, becomes
\ba
S[\,g\,] = &-& \frac{\Lambda}{2\kappa{\cal G}} \int d^5X
\sqrt{g(X)}\left\{ R(X)\ -\ 2\lambda\right\}\no
\simeq &-& \frac{\Lambda}{2\kappa{\cal G}} \int d^4 x \sqrt{g(x)}\, R(x)
\int_{-\infty}^{+\infty}dz\ \exp\{-2\rho(z)\}\no
&-& \frac{\Lambda}{\kappa{\cal G}} \int d^4 x \sqrt{g(x)}
\int_{-\infty}^{+\infty}dz\ \exp\{-4\rho(z)\}\left\lbrace
6[\,\rho^\prime(z)\,]^2 - \lambda\right\rbrace\no
\equiv &-& \frac{1}{16\pi G_N} \int d^4 x \sqrt{g(x)}\left\lbrace
R(x)\ -\ 2\Lambda_{\rm grav}\right\rbrace\ ,
\label{fivefour}
\ea
whence we eventually get the Planck mass 
scale $M_{P} \sim 1.22\times 10^{19}$ GeV/c$^2$ which
corresponds to the Newton's gravitational constant
\be
M_P^2 = G_N^{-1} \equiv \frac{8\pi\Lambda}{\kappa{\cal G}}
\int_{-\infty}^{+\infty}dz\ \exp\{-2\rho(z)\}
\ee
and the gravitational part of the four dimensional cosmological constant
\be
\Lambda_{\rm grav}\equiv
\frac{8\pi G_N\Lambda}{\kappa{\cal G}}
\int_{-\infty}^{+\infty}dz\ \exp\{-4\rho(z)\}\left\lbrace
\lambda - 6[\,\rho^\prime(z)\,]^2\right\rbrace\ .
\ee
Under the assumption of the smallness of the parameter ${\overline \kappa} \ll 1$
and using in eq.~\gl{fivefour} the approximate form for $|z|\to\infty$
of the solution \gl{rhoone} one obtains
\ba
G_N\ \simeq\ \frac{\pi^2\,{\overline \kappa}^{\,2}}{2N\Lambda M}\ .
\ea
Remarkably, the full value of the  cosmological constant, including the gravitational
as well as the matter vacuum energy densities, does indeed exactly vanish 
to all orders in the perturbative expansion in powers of 
${\overline \kappa}\,:$ actually,
\ba
\Lambda_{\rm cosmo}&\equiv& 
\frac{2N\Lambda G_N}{\pi^2}
\int_{-\infty}^{+\infty}dz\ \exp\{-4\rho(z)\}\biggl\lbrace
2M^2 \lambda_{\rm eff}  - ({4/{\overline \kappa}})\,{M^2}
\rho^{\prime\,2}(z)\no
&+& \Phi^{\prime\,2}(z) + H^{\prime\,2}(z) - 2\Delta_1\Phi^2(z) - 2\Delta_2 H^2(z)
+ \left[\,\Phi^2(z) + H^2(z)\,\right]^2\no
&+& \frac23\left[\,\Phi^2(z) + H^2(z)\,\right]
\left[\,2\rho^{\prime\prime}(z) - 5\rho^{\prime\,2}(z)\,\right]\biggr\rbrace\no
&=& \frac{4N\Lambda G_N}{\pi^2{\overline \kappa}}
\int_{-\infty}^{+\infty}dz\ \exp\{-4\rho(z)\}\biggl\lbrace
-\ M^2\rho^{\prime\prime}(z) + {\overline \kappa}\left[\,
\Phi^{\prime\,2}(z) + H^{\prime\,2}(z)\,\right]\no
&+& \frac23\,{\overline \kappa}\left[\,\Phi^2(z) + H^2(z)\,\right]
\left[\,2\rho^{\prime\prime}(z) - 5\rho^{\prime\,2}(z)\,\right]\biggr\rbrace\
=\ 0\ , 
\label{zerocos}
\ea
where the vacuum expectation values \gl{eqscalar1} and \gl{eqscalar2}
of the scalar fields together with the field equation \gl{eqgrav05}
of the conformal factor have been suitably taken into account.
In order to establish the exact cancellation between the gravitational 
and the scalar matter contributions, the five dimensional
cosmological constant has been conveniently substituted from eq.~\gl{intcosm}.
We also notice that the inclusion of the higher-order gravitational interaction
in the last line of eq.~\gl{confm} still keeps equal to zero the value
of the induced cosmological constant $\Lambda_{\rm cosmo}\,,$ 
as it can be shown with the help of the Appendix C.

Let us find the relations among the AdS curvature scale
$k\simeq 2{\overline \kappa}\,M/3$, the Planck
mass $M_P$ and the spontaneous symmetry breaking scale $M\,.$
We recall that the characteristic parameter of ultra-low energy 
dynamics of light fermions and scalar fields has been found 
\cite{aags} to be 
\be
\zeta \equiv \frac{M\pi^3}{N \Lambda}\ ,
\ee
with $N \sim \pi^3$ in the Standard Model (see the summary in the previous Section 4).
Therefore, the relationship among the three scales $M_{P}, k $ and $M$ actually reads
\be
k^2 M_{P}^2 = \frac{8\pi}{9\zeta}\,M^4\ ,
\label{scales}
\ee
in accordance with Eqs.~\gl{newcoup} and \gl{AdSlim}. 
Correspondingly one obtains that
\be
\bar\kappa = \frac{2\pi}{\sqrt{\zeta}}\,(M/M_P) 
\la{barkap}  
\ee
restricted to $\bar\kappa \ll 1$.

One can now express both the five dimensional gravitational constant and
the five dimensional Planck scale, respectively, just in terms of
the above mentioned scales and parameters, that means
\be
\frac{{\cal G}\kappa}{\Lambda} = \frac{6\pi^3{\overline \kappa}}{N\Lambda M^2}
= \frac{9 k\zeta}{M^4}\,=\,\frac{1}{M_*^3}\ ;\quad M_*^{3} = \frac{k M^2_P}{8\pi}\ ,
\la{planck5}
\ee
which leads to the lower bound  
for the five dimensional Planck scale $ M_* > 10^8$ GeV
from the experimental bound $ k > 10\
\mbox{\rm mm}^{-1} = 2\cdot 10^{-3}$ eV, the latter one being based on possible deviations
from the Newton's law \ci{adelb}.

Our first scenario -- {\tt fundamental gravity} -- 
is selected to have a principally detectable dynamics of 
scalar fields and of the Higgs-Yukawa
coupling to fermions in the brane world. 
Let us therefore analyze the regime where
$M/\Lambda \simeq \zeta \sim 0.1 \div 0.3 \,,$
at least, albeit not much less. 
In this case, the experimental bound $ k > 10\
\mbox{\rm mm}^{-1} = 2\cdot 10^{-3}$ eV \ci{adelb} turns out to be 
compatible with the lower bound \ci{rev3,rev9} for the localization scale
$M > (2 \div 3)$ TeV, which makes somewhat 
challenging to produce new physics related
to the fifth dimension at the next generation of colliders. 
The corresponding cut-off is $\Lambda > 10\div 20$ TeV so that,
from eq.~\gl{barkap}, perturbation theory is controlled by
the very tiny constant ${\overline \kappa} > 10^{-15}\,,$ 
whilst the five dimensional cosmological 
constant must be tuned to the value
$\lambda \sim 10^{-2}$ MeV$^2\,.$  

Let us find the relationship between the bare value ${\cal G}$ and
dressed value $\kappa {\cal G}$ of the five dimensional gravitational constants. 
According to eq.~\gl{kappa}, the latter one is controlled by the ratio
\be
\omega =  \frac{N \Lambda^2 {\cal G}}{54\pi^3} = 
\frac{{\overline \kappa}\,\pi^6}{9 N^2 \zeta^2} \sim 10\, 
{\overline \kappa} \sim 10^{-14}\ll 1 
\la{omega} 
\ee
for the chosen values of parameters ${\overline \kappa}$ and $\zeta\,.$  
Thus we conclude that $\kappa\simeq 1$ and therefore the bare gravitational 
constant ${\cal G}$ mostly determines the intensity of the gravitational 
attraction in the five dimensional space-time.
But this gravitational force is pretty strong, as its Planck scale 
$M_*  \sim 10^8$ GeV is much lower than its four dimensional counterpart
$M_P \sim 10^{19}$ GeV.

A further economical choice might be to identify the AdS curvature
scale  $ k$ with the electroweak symmetry breaking scale $\mu \sim 200$ GeV 
with the hope to connect  the top-quark mass formation in the Standard Model 
to some extra dimensional gravitational effects.
If $ k \sim \mu  \sim 200$ GeV, and still $\zeta \sim 0.1 \,$ 
then one finds $M \sim  10^{10}$ GeV and $\Lambda \sim 10^{11}$ GeV so that
the five dimensional cosmological constant must be tuned to $\lambda\sim 10^{12}$ GeV$^2$.

It follows therefrom that the expansion parameter increases up to
${\overline \kappa} \sim 10^{-8}\,.$  Nevertheless,
it does keep to be very small and, consequently, one basically needs the first orders of
perturbation theory in the gravitational interaction to the aim of reaching
a sufficiently good precision. For such a choice, the only signatures of
extra dimensional physics could come from branon detection and
there is no hope to reach sufficiently high energies to overcome 
the barrier $M$ towards the fifth dimension. 
The estimates coming from eqs.~\gl{omega} and \gl{planck5}, {\it i.e.}
$\omega \sim 10^{-8}\ll 1\,;\quad M_{*} \sim 10^{13}$ GeV,
show that this scenario still corresponds to the fundamental
five dimensional gravity Action with $\kappa \simeq 1\,,$ 
a relatively strong gravity.  

The second scenario under consideration is that one of the 
{\tt induced gravity}. Let us adopt the induced gravity relations
of eq.~\gl{indgrav} 
\be 
\kappa\ \simeq\ {54\pi^3}/{N\Lambda^2 {\cal G}}\ ;\qquad 
{\overline \kappa}^{\,1/2}\ \simeq\ {3M/\Lambda}\ =\ 
{3 N\zeta}/{\pi^3} \ll 1\ ,
\ee
in such a way that
\be 
k M_{P}^2 = {4N\Lambda^3}/{27\pi^3}\ ;\qquad 
k^5 M_{P}^4 = {128N^2 M^9}/{27\pi^6}\ .
\ee
In particular, for a lower experimental bound 
$k \geq 2\cdot 10^{-12}$ GeV one finds 
\ba
&& M \geq 100\ {\rm GeV}\ ,\qquad
\Lambda \geq 10^{9}\ {\rm GeV}\ ;\no
&& \zeta \sim M/\Lambda \sim 10^{-7}\ ,\qquad 
{\overline \kappa} \sim 10^{-13}\ .
\ea

This means that the light particle interaction is highly suppressed and
the only particle interaction which is left is the gravitational one.
It is worthwhile to remark that the induced gravity still keeps to be weak 
at low energies 
${\overline \kappa} \ll 1\,,$ which also entails the equivalence between 
the  compositeness scale $\Lambda_c$ defined in eq.~\gl{compo} and 
the five dimensional Planck-like scale $M_{*}> 10^8$ GeV (see
eq.~\gl{planck5}).  Evidently the barrier $M \sim 100$ GeV 
is too low to be accepted by modern collider experiments, as the
gravity (and gauge bosons) may easily be able to give fermions enough
energy to disappear from our world.
 
For so far experimentally acceptable barriers of order
$M \sim 1$ TeV, one finds that the AdS curvature scale 
$k \sim  10^{-10}$ GeV, which corresponds to distances 
of the ${\mu}$m order, becomes  certainly unreachable in a
nearest future experiment hunting for Newton's law deviations 
\ci{adelb}. As a matter of fact, the scalar particles essentially decouple
from the fermion world and from each other since
$\Lambda_c \simeq M_{*} \sim 10^{9}$ GeV and 
$\zeta \sim M/\Lambda \sim 10^{-6.5}\,;\ {\overline \kappa} \sim 10^{-12}\,.$
Although the Higgs-like particles may be involved into the gauge boson 
interaction and be observable by gauge boson mediation, it turns out
that branons, {\it i.e.} the quanta of the field $\phi$, 
do represent excitations of a geometrical nature, being related to
the Goldstone bosons of the translational invariance breaking. 
As a consequence, their actual decoupling
from any other kind of matter makes them a perfect candidate for the dark
matter/energy, depending on their mass.

For the electroweak breaking scale $k \sim 2\cdot 100$ GeV,
the barrier $M \sim 10^{10}$ GeV is too high to trigger any observable
physics at the Earth laboratories since
$\Lambda \sim 10^{14} {\rm GeV}\,;\ \zeta \sim M/\Lambda \sim 10^{-4}\,,$
whereas branons keep essentially decoupled and thereby belonging to the dark
universe. The low-energy gravity remains weak with 
${\overline \kappa} \sim 10^{-8}$ so that
we see that, for matter induced gravity, the parameter 
$\zeta$ may not be small only for scales $k \sim  M \sim \Lambda$ 
approaching the Planck scale, where gravity is actually strong 
(${\overline \kappa}\sim 1$) and quantum gravity is in order.


\section{Conclusions, further development}

In this paper we have performed the systematic analysis of how the 
brane world can be generated dynamically by matter self-interaction 
from a spontaneous breaking of the translational
invariance. The latter breaking is intimately related to the
$\tau$-symmetry breaking in our proto-fermion model. 
The pertinent low-energy effective Action for fermions and the
auxiliary scalar fields, as well as for
their gravitational interaction, has been obtained by one-loop 
integration of high energy spinor degrees of freedom. 

Both the gravitational and the scalar fields turn out to be
responsible for the localization of light matter on the
brane. Nevertheless, even if the scalar kink-like vacuum solution 
allows to trap light fermion and scalar particles inside the
brane layer, still the AdS$_5$ gravitational background, in which 
the kink-like vacuum configuration are embedded, just induces 
the quantum tunneling taking away, in principle, some of those 
particles. Meanwhile, some simple WKB estimations prove that such a decay,
eventually due to the quantum tunnel effect, 
is extremely well suppressed, so that it becomes practically impossible
to detect the disappearance of any particle in the visible universe during
the universe lifetime. 

These circumstances have allowed us to formulate the following principle:\\
{\sl particle lifetimes on the brane must be evaluated by wave packet localization, 
which is unambiguously and consistently determined in perturbation theory.} 
Within this approach, the mass spectrum of light states on the brane turns out
to be slightly different from the corresponding one in the flat space limit.

We have in turn examined the possible ranges of different scales and coupling
constants in terms of the Newton's constant normalization to its observed value
and in terms of the experimental bounds on the AdS$_5$ curvature and thresholds
on appearance/disappearance of high energy particles in accelerator experiments. 
As a summary of these studies, we conclude that:\\
a) on the one hand, fundamental gravity in five dimensions appears to be  
more challenging 
for future experiments, because the phenomenologically acceptable large values 
of AdS$_5$ curvature $k \sim 10^{-3}$ eV together with the relatively low 
translational symmetry breaking scales $M \sim 1 \div 3$ TeV turn out to be
compatible with  a weak, but not vanishing, coupling of branons to Higgs-like 
scalars and fermions. However, it appears that the lower are the values 
of the AdS$_5$ curvature $k$, the higher is the threshold for new physics $M$ 
and the weaker is the interaction among spinor and scalar matter, in such a
way to move branons to the dark side of the universe.\\
b) On the other hand, induced gravity leads to decoupling of branons 
from other matter in a wide range of acceptable scales and coupling
constants, thus putting them straightforwardly to the dark matter realm.\\
c) In any case, the dimensionless parameter which characterizes the
strength of the gravitational interaction is very small, of the order 
${\overline \kappa} \leq 10^{-8}$. This feature does justify the use of
the perturbation theory both in the calculation of vacuum field 
configurations and gravity background, as well as in the derivation of
mass spectrum of localized particles.

For nearest future developments, we deserve the following interesting 
problems:\\
a) interplay in deviation from the scaling point (where only zero-mass
states are localized on the brane) between 
manifest driving with the scale $\mu$
and mass generation by the AdS background gravity; \\
b) next-to-leading effects in the gravitational coupling constant
$\bar\kappa$ 
and, in
particular, scalar gravity (``radion''\ci{rad1,rad2,rad3}) influence
on the matter spectrum;\\
c) soft breaking of the translational symmetry due to scalar field defects, 
necessarily accompanied with defects in the cosmological constant and
consequent production of massive branons;\\
d) supplementing our model with gauge bosons;\\
e) search for new experimental perspectives to detect branons and
other signatures of extra dimensions based on the model presented in
our paper.

\acknowledgments
We are grateful to Luca Vecchi for his accurate check of certain formulas.
This work was supported by Grant INFN/IS-PI13. Two of us (A.A.A. and V.A.A.)
were supported by Grants RFBR 04-02-26939 and Grant UR.02.01.299.
\appendix
\section{One-loop effective Action}

Let us consider the second order matrix-valued positive
elliptic differential operator
\ba
\ddi^\dagger\ddi &=&\ -\ g^{AB}(X)D_A D_B +
{\cal M}^2(X)\ , \no
{\cal M}^2(X) &\equiv& \frac14\,R(X) +
\Phi^2(X)  + H^2(X)   - \tau_3 \not\!\partial \Phi(X)
 - \tau_1 \not\!\partial H(X)\ ,\la{quad1}
\ea
acting on a $n$-dimensional Riemannian manifold (eventually $n = 5$).
Our aim is to evaluate the matrix element of the distribution of the states
operator: namely,
\be
\langle X|\vartheta(Q^2 - \ddi^\dagger\ddi)|Y\rangle =
\int_{c-i\infty}^{c+i\infty}\frac{d t}{2\pi i}\ \frac{\exp\{tQ^2\}}{t}\,
\langle X|\exp\{-t\ddi^\dagger\ddi\}|Y\rangle\ ,\quad c>0\ .\la{a1}
\ee
As a matter of fact, if the positive operator $\ddi^\dagger\ddi$
is also of the trace class,
then the trace of $\vartheta(Q^2 -\ddi^\dagger\ddi )$ does represent the number
of the eigenstates of $\ddi^\dagger\ddi$ up to the momentum square $Q^2$.
It is convenient to write the heat kernel in the form
\be
\langle X|\exp\{-t\ddi^\dagger\ddi\}|Y\rangle\equiv
(4\pi t)^{-n/2}\exp\left\{-\frac{(X-Y)^2}{4t}\right\}
\Omega(t|X,Y)\ ,\la{a2}
\ee
where $\Omega(t|X,Y)$ is the so called transport function
which fulfills
\ba
&&\left(\partial_t+\frac{X\cdot\partial}{t}+
\ddi^\dagger_X\ddi_X\right)\Omega(t|X,Y)=0\ ,\\
&&\lim_{t\downarrow 0}\Omega(t|X,Y)=\widehat{\bf 1}\ ,\la{a3}
\ea
in such a way that
\be
\lim_{t\downarrow 0}\ \langle X|\exp\{-t\ddi^\dagger\ddi\}|Y\rangle=
\widehat{\bf 1}\delta^{(n)}(X-Y)\ .\la{a4}
\ee
If we insert eq.\,\gl{a2} in eq.\,\gl{a1} and change the integration variable
we obtain
\ba
\langle X|\vartheta(Q^2 - \ddi^\dagger\ddi)|Y\rangle&=&
2Q^n (4\pi)^{-1-n/2}\int_{c-i\infty}^{c+i\infty}
dt\ t^{-1-n/2}\no
&\times&\exp\left\{t-\frac{Q^2(X-Y)^2}{4t}-\frac{i\pi}{2}\right\}
\Omega\left(\frac{t}{Q^2}\left|X,Y\right.\right)\ .\la{a5}
\ea
Now, if we write
\be
\Omega\left(\frac{t}{Q^2}\left|X,Y\right.\right)=
\sum_{k=0}^{[n/2]}t^k a_k(X,Y)Q^{-2k}
+{\cal R}_{[n/2]+1}\left(\frac{t}{Q^2}\left|X,Y\right.\right)\ ,\la{a6}
\ee
it turns out that, by insertion of eq.\,\gl{a6} into the integral
\gl{a5}, the last term in
the RHS of the above expression becomes sub-leading and negligible
in the limit of very large $Q$.
As a consequence, the leading asymptotic behavior of the matrix
element \gl{a1} in the large $Q$ limit reads \cite{GR}
\ba
\langle X|\vartheta(Q^2 - \ddi^\dagger\ddi)|Y\rangle
&\stackrel{Q\rightarrow\infty}{\sim}&
(4\pi)^{-n/2}\sum_{k=0}^{[n/2]}a_k(X,Y) Q^{n-2k}\no
&\times&\int_{c-i\infty}^{c+i\infty}
\frac{dt}{2\pi i}\ t^{k-1-n/2}\,\exp\left\{t-\frac{Q^2(X-Y)^2}{4t}\right\}\no
&\stackrel{X\rightarrow Y}{\sim}&
(4\pi)^{-n/2}\sum_{k=0}^{[n/2]}a_k(X,Y)\,
\frac{Q^{n-2k}}{\Gamma(1-k+n/2)}\ .\la{a7}
\ea
In the diagonal limit $X=Y$, for $n=4,5$ the relevant
coefficients of the heat kernel
asymptotic expansion take the form \ci{cogia}
\ba
a_0(X,X)&\equiv& \widehat{\bf 1}\ ;\qquad
a_1(X,X)={R(X)\over 6} - {\cal M}^2(X)\ ;\no
a_2(X,X)&=&\frac{1}{2}[\,a_1(X,X)\,]^2
-\frac{1}{6}\,D^2{\cal M}^2(X)
+\frac{1}{30}\,D^2 R(X)\no
&+&\frac{1}{180}\left[\,
R^{ABCD}(X)R_{ABCD}(X) - R^{AB}(X)R_{AB}(X)\,\right]\no
&+&\frac{1}{12}\,\Omega^{AB}(X)\Omega_{AB}(X)
\ ,\la{a8}
\ea
where the curvature bundle, {\it i.e.} the field strength associated to
the spin connection, reads
\be
\Omega_{AB}(X)\equiv \frac18\,\left[\,\widehat\gamma_j,\widehat\gamma_k\,\right]\,
R_{ABjk}(X)\ .
\ee
In such a way, we can write down the dominant diagonal matrix element,
leading eventually to the five dimensional Euclidean
low--energy Lagrange density, in the form
\ba
&&\langle X|\vartheta(Q^2 - D^\dagger D)|X\rangle
\stackrel{Q\rightarrow\infty}{\sim}\no
&&\frac{Q^5}{60\pi^3}\left\{
\widehat{\bf 1}+\frac{5}{2}\,Q^{-2}\,a_1(X,X)+
\frac{15}{4}\,Q^{-4}\,a_2(X,X)\right\}\ .
\ea
Furthermore we find \quad[ ${\rm tr}\,a_0(X,X)={\rm tr}\,\widehat{\bf 1} = 8$ ]
\ba
\frac18\,{\rm tr}\,a_1(X,X)&=&
- \frac{1}{12}\,R(X) - [\,\Phi^2(X) + H^2(X)\,]\ ;\\
\frac18\,{\rm tr}\,[\,a_1(X,X)\,]^2 &=&
\frac{1}{144}\,R^2(X) + \frac{1}{6}\,R(X)[\,\Phi^2(X) + H^2(X)\,]\no
&+&\partial_A\Phi(X)\partial^A\Phi(X) +
\partial_A H(X)\partial^A H(X)\no
&+& [\,\Phi^2(X) + H^2(X)\,]^2
\ea
so that we eventually obtain
\ba
\frac18\,{\rm tr}\,a_2(X,X)&=&
\frac12\,\partial_A\Phi(X)\partial^A\Phi(X) +
\frac12\,\partial_A H(X)\partial^A H(X)\no
&+&\frac12[\,\Phi^2(X) + H^2(X)\,]^2 +
\frac{1}{12}\,R(X)[\,\Phi^2(X) + H^2(X)\,]\no
&-&\frac16\,D^2[\,\Phi^2(X) + H^2(X)\,]-\frac{1}{120}\,D^2 R(X)
+ \frac{1}{288}\,R^2(X)\no
&-&\frac{1}{180}R^{AB}(X)R_{AB}(X)
-\frac{7}{1440}\,R^{ABCD}(X)R_{ABCD}(X)\ .
\ea

\section{Curvature tensors for conformal metric}

Consider the five dimensional Riemannian metric
\ba
&&g_{\mu\nu}(x,z)=g_{\mu\nu}(x)\,\exp\{-2\rho(z)\}\ ,\no
&&g_{\mu 5}(x,z)\ =\ g_{5\nu}(x,z)\ =\ 0\ ,\no
&&g_{55}(x,z)\ =\ 1\ .
\ea
The related Christoffel symbols take the values
\ba
&&\Gamma^\lambda_{\mu\nu}(x,z)=
\frac12\,g^{\lambda\kappa}(x)\left\lbrace
\partial_\mu g_{\nu\kappa}(x) + \partial_\nu g_{\mu\kappa}(x)
- \partial_\kappa g_{\mu\nu}(x)\right\rbrace
\equiv \Gamma^\lambda_{\mu\nu}(x)\ ,\no
&&\Gamma^\lambda_{\mu 5}(x,z)\ =\ \Gamma^\lambda_{5\mu}(x,z)\ =\
-\ \rho^\prime(z)\,\delta^\lambda_\mu\ \equiv\ \Gamma^\lambda_{\mu 5}(z)\ ,\no
&&\Gamma^5_{\mu\nu}(x,z)\ =\ \rho^\prime(z)\,
\exp\{-2\rho(z)\}\,g_{\mu\nu}(x)\ ,\no
&&\Gamma^5_{5\mu}(x,z)\ =\ \Gamma^\lambda_{55}(x,z)\ =\
\Gamma^5_{55}(x,z)\ =\ 0\ .
\ea
The corresponding Riemann tensor has components
\ba
R^\alpha_{\ \beta\mu\nu}(x,z) &=&
R^\alpha_{\ \beta\mu\nu}(x)\ +\
[\,\rho^\prime(z)\,]^2\,\exp\{-2\rho(z)\}\,
\left\{g_{\beta\mu}(x)\,\delta_\nu^\alpha -
g_{\beta\nu}(x)\,\delta_\mu^\alpha\right\}\ ,\no
R_{5\beta 5\nu}(x,z)\ &=&\ \exp\{-2\rho(z)\}\,g_{\beta\nu}(x)
\left\{\rho^{\prime\prime}(z) - [\,\rho^\prime(z)\,]^2\right\}\ ,\no
R^\alpha_{\ 5\mu\nu}(x,z)\ &=&\ R^\alpha_{\ \beta 5\nu}(x,z)\ =\
R_{5\beta\mu\nu}(x,z)\ =\ 0\ .
\ea
The components of the Ricci tensor read
\ba
R_{\beta\nu}(x,z) &=&
R_{\beta\nu}(x) + \exp\{-2\rho(z)\}\,g_{\beta\nu}(x)
\left\{\rho^{\prime\prime}(z) - 4[\,\rho^\prime(z)\,]^2\right\}\ ,\no
R_{55}(x,z)\ &=& 4\left\{\rho^{\prime\prime}(z) - [\,\rho^\prime(z)\,]^2\right\}\ ,\no
R_{\beta 5}(x,z) &=& 0\ ,
\ea
so that the scalar curvature becomes
\be
R(x,z) = \exp\{2\rho(z)\}\,R(x) + 8\rho^{\prime\prime}(z)
-20[\,\rho^\prime(z)\,]^2\ .
\label{scalarcur5}
\ee
It is also useful to compute the quadratic invariant in the Riemann tensor:
we find
\ba
&&R_{ABCD}(x,z)\,R^{ABCD}(x,z) =\no
&&R_{\alpha\beta\mu\nu}(x,z)\,R^{\alpha\beta\mu\nu}(x,z) +
4R_{\alpha 5\mu 5}(x,z)R^{\alpha 5\mu 5}(x,z)\ .
\ea
We have
\ba
R^{\alpha\beta\mu\nu}(x,z) &=&
\exp\{6\rho(z)\}\,R^{\alpha\beta\mu\nu}(x)\no
&+&
[\,\rho^\prime(z)\,]^2\exp\{4\rho(z)\}\,
\left\{g^{\alpha\nu}(x)g^{\beta\mu}(x) -
g^{\alpha\mu}(x)g^{\beta\nu}(x)\right\}\ ,\no
R_{\alpha\beta\mu\nu}(x,z) &=&
\exp\{-2\rho(z)\}\,R_{\alpha\beta\mu\nu}(x)\no
&+&
[\,\rho^\prime(z)\,]^2\exp\{-4\rho(z)\}\,
\left\{g_{\alpha\nu}(x)g_{\beta\mu}(x) -
g_{\alpha\mu}(x)g_{\beta\nu}(x)\right\}\ ,
\ea
so that
\ba
R_{\alpha\beta\mu\nu}(x,z)\,R^{\alpha\beta\mu\nu}(x,z)&=&
\exp\{4\rho(z)\}\,R_{\alpha\beta\mu\nu}(x)\,R^{\alpha\beta\mu\nu}(x)\no
&-& 4\exp\{2\rho(z)\}\,[\,\rho^\prime(z)\,]^2 R(x)
+24[\,\rho^\prime(z)\,]^4
\ea
and finally
\ba
R_{ABCD}(x,z)\,R^{ABCD}(x,z) &=&
16[\,\rho^{\prime\prime}(z)\,]^2 -
32\rho^{\prime\prime}(z)[\,\rho^\prime(z)\,]^2 +
40[\,\rho^\prime(z)\,]^4\no
&+&\exp\{4\rho(z)\}\,R_{\alpha\beta\mu\nu}(x)\,R^{\alpha\beta\mu\nu}(x)\no
&-& 4\exp\{2\rho(z)\}\,[\,\rho^\prime(z)\,]^2 R(x)\ .
\ea
In a similar way we obtain the quadratic invariant in the Ricci tensor
that reads
\ba
R_{AB}(x,z)\,R^{AB}(x,z) &=&
R_{\alpha\beta}(x,z)\,R^{\alpha\beta}(x,z) + R_{55}(x,z) R^{55}(x,z)\no
&=& 20[\,\rho^{\prime\prime}(z)\,]^2 -
64\rho^{\prime\prime}(z)[\,\rho^\prime(z)\,]^2 +
80[\,\rho^\prime(z)\,]^4\no
&+& 2\exp\{2\rho(z)\}\,R(x)\left\{\rho^{\prime\prime}(z)
-4[\,\rho^\prime(z)\,]^2\right\}\no
&+&\exp\{4\rho(z)\}\,R_{\alpha\beta}(x)\,R^{\alpha\beta}(x)\ .
\ea
Finally, we easily get the quadratic invariant in the curvature scalar
\ba
R^2(x,z) &=& 64[\,\rho^{\prime\prime}(z)\,]^2
+ 400[\,\rho^\prime(z)\,]^4
- 320\rho^{\prime\prime}(z)[\,\rho^\prime(z)\,]^2\no
&+&
8\exp\{2\rho(z)\}\,R(x)\left\{2\rho^{\prime\prime}(z)
- 5[\,\rho^\prime(z)\,]^2\right\}\no
&+& \exp\{4\rho(z)\}\,R^2(x)\ .
\ea
\bigskip
Consider now the special case of a {\sl quasi--flat} Riemannian metric
\ba
&&g_{\mu\nu}(z)=\delta_{\mu\nu}\,\exp\{-2\rho(z)\}\ ,\no
&&g_{\mu 5}(z)\ =\ g_{5\nu}(z)\ =\ 0\ ,\no
&&g_{55}(z)\ =\ 1\ .
\ea
The related Christoffel symbols take the values
\ba
&&\Gamma^\lambda_{\mu\nu}(z)\ =\ 0\ ,\no
&&\Gamma^\lambda_{\mu 5}(z)\ =\ \Gamma^\lambda_{5\mu}(z)\ =\
-\ \rho^\prime(z)\,\delta^\lambda_\mu\ ,\no
&&\Gamma^5_{\mu\nu}(z)\ =\ \rho^\prime(z)\,
\exp\{-2\rho(z)\}\,\delta_{\mu\nu}\ ,\no
&&\Gamma^5_{5\mu}(z)\ =\ \Gamma^\lambda_{55}(z)\ =\
\Gamma^5_{55}(z)\ =\ 0\ .
\ea
where $\rho^\prime(z)\equiv(d\rho/dz)\,.$
The corresponding Riemann tensor has components
\ba
R^\alpha_{\ \beta\mu\nu}(z) &=&
[\,\rho^\prime(z)\,]^2\,\exp\{-2\rho(z)\}\,
\left\{\delta_{\beta\mu}\,\delta_\nu^\alpha -
\delta_{\beta\nu}\,\delta_\mu^\alpha\right\}\ ,\no
R_{5\beta 5\nu}(z)\ &=&\ \exp\{-2\rho(z)\}\,\delta_{\beta\nu}
\left\{\rho^{\prime\prime}(z) - [\,\rho^\prime(z)\,]^2\right\}\ ,\no
R^\alpha_{\ 5\mu\nu}(z)\ &=&\ R^\alpha_{\ \beta 5\nu}(z)\ =\
R_{5\beta\mu\nu}(z)\ =\ 0\ .
\ea
The components of the Ricci tensor read
\ba
R_{\beta\nu}(z) &=&
\exp\{-2\rho(z)\}\,\delta_{\beta\nu}
\left\{\rho^{\prime\prime}(z) - 4[\,\rho^\prime(z)\,]^2\right\}\ ,\no
R_{55}(z)\ &=& 4\left\{\rho^{\prime\prime}(z) - [\,\rho^\prime(z)\,]^2\right\}\ ,\no
R_{\beta 5}(z) &=& 0\ ,
\ea
so that the scalar curvature becomes
\be
R(z) = 8\rho^{\prime\prime}(z)
-20[\,\rho^\prime(z)\,]^2\ .
\ee
The corresponding quadratic invariants in the Riemann tensor, the Ricci tensor
and the curvature scalar become
\ba
&&R_{ABCD}(z)\,R^{ABCD}(z) =
16[\,\rho^{\prime\prime}(z)\,]^2 -
32\rho^{\prime\prime}(z)[\,\rho^\prime(z)\,]^2 +
40[\,\rho^\prime(z)\,]^4\\
&&R_{AB}(z)\,R^{AB}(z) =
20[\,\rho^{\prime\prime}(z)\,]^2 -
64\rho^{\prime\prime}(z)[\,\rho^\prime(z)\,]^2 +
80[\,\rho^\prime(z)\,]^4\\
&&R^2(z) = 64[\,\rho^{\prime\prime}(z)\,]^2
- 320\rho^{\prime\prime}(z)[\,\rho^\prime(z)\,]^2
+ 400[\,\rho^\prime(z)\,]^4\ .
\ea
Finally, for a scalar function $f(z)$ of the fifth coordinate we have
\ba
&&D^C\partial_C f(z)\ =\ f^{\,\prime\prime}(z) - 4\rho^{\,\prime}(z)\,f^{\,\prime}(z)\ ;\\
&&\left\{g_{\alpha\alpha}(z)D^C\partial_C - D_\alpha\partial_\alpha\right\}f(z)\ =\no
&&\exp\{-2\rho(z)\}\left\{f^{\,\prime\prime}(z) -
3\rho^{\,\prime}(z)\,f^{\,\prime}(z)\right\}\ ;\\
&&\left(D^C\partial_C - D_5\partial_5\right)f(z)\ =\
- 4\rho^{\,\prime}(z)\,f^{\,\prime}(z)\ .
\ea


\section{Equations of motion in conformal metric}

In the case of conformally flat metric \gl{metr} and for 
a kink-like pair of scalar fields
\be
\langle\Phi(X)\rangle_0 = \Phi(z)\ ,\qquad 
\langle H(X)\rangle_0 = H(z)\ ,
\ee 
from the full low-energy Lagrange density (\ref{lowmin}) 
one finds the following gravitational field equations in
which the quadratic terms in the curvature tensors are suitably
taken into account: namely,
\ba
&& \rho^{\prime\prime}(z)
\left(\delta_{A5}\,\delta_{B5} - 1\right) 
+ \left[\,2\rho^{\prime\,2}(z)  
+ {\lambda}/{3}\,\right]\delta_{AB}\ =\no  
&& =\ ({N \kappa\cal{G}}/{12\pi^3})\left\lbrace 
\left(2 \delta_{A5}\,\delta_{B5} - \delta_{AB}\right)
\left[\,\Phi^{\prime\,2}(z) + H^{\prime\,2}(z)\,\right]\right.\no
&& +\,\ \delta_{AB}\left\lgroup 2\Delta_1\Phi^2(z) + 2\Delta_2 H^2(z) -
\left[\,\Phi^2(z) + H^2(z)\,\right]^2\right\rgroup\no
&& +\ \left[\, \rho^{\prime\prime}(z)
\left(\delta_{A5}\,\delta_{B5} - \delta_{AB}\right)
+ 2\rho^{\prime\,2}(z)\delta_{AB}\,\right]
\left[\,\Phi^{\prime\,2}(z) + H^{\prime\,2}(z)\,\right]\no
&& +\,\ (1/3)\left\lgroup \delta_{AB}
\left[\,\partial_z^2 - 3\rho^\prime(z)\,\partial_z\,\right] 
- \delta_{A5}\,\delta_{B5}\left[\,\partial_z^2 
+ \rho^\prime(z)\,\partial_z\,\right]\right\rgroup\ \times\no
&&\times\ \left[\,\Phi^{\prime\,2}(z) + H^{\prime\,2}(z)\,\right]
-\ \left.\delta_{AB}\ F_1(R^2)\ +\ 
\delta_{A5}\,\delta_{B5}\ F_2(R^2)\right\rbrace\ ,
\label{eqgrav3}
\ea
supplemented with eqs.~\gl{eqscalar1} for matter fields.
Here the higher order terms $F_{1,2} (R^2)$ can be calculated from 
that part of the low-energy Euclidean Action \gl{lowmin},
which is quadratic in the curvature tensors, as we shall see below.
These equations can be derived from the effective Action with 
Lagrange density \gl{confm} by means of two types of variations. 
The variation with respect to the conformal factor $\rho(z)$ gives rise to 
the contribution into the partial trace of eq.s~\gl{eqgrav3}, whereas 
the infinitesimal change of the $z$-coordinate,
$dz^\prime = dz[\,1 - \epsilon(z)\,]$ drives to the variation of the 
$g_{55}(z)$ component of the metric \gl{metr}. 
The latter equation can be derived directly from eq.s~\gl{confm} 
according to the following rules: namely,
\ba
\delta g_{55}(z)\ &=& -\ 2\epsilon(z)\ ;\qquad 
\delta\rho^\prime(z)\ =\ \rho^\prime(z)\,\epsilon(z)\ ;\no
\delta\rho^{\prime\prime}(z)\ &=& 
2\rho^{\prime\prime}(z)\,\epsilon(z)\ 
+\ \rho^\prime(z)\,\epsilon^\prime(z)\ .
\ea 
Now, from the equality  
\begin{displaymath}
\frac{N\Lambda}{2880\pi^3}\int d^5X\sqrt{g}
\left\lbrace 5R^2(X) -
8R_{AB}(X)R^{AB}(X) - 7R_{ABCD}(X)R^{ABCD}(X)\right\rbrace
\end{displaymath}
\begin{displaymath}
=\ \frac{N\Lambda}{120\pi^3}\int d^5X\ \exp\{-4\rho(z)\}\left\lbrace\,
2[\,\rho''(z)\,]^2  - 36 \rho''(z)[\,\rho'(z)\,]^2 
+ {45}[\,\rho'(z)\,]^4\,\right\rbrace
\la{squared}
\end{displaymath}
and by means of the above mentioned two kinds of variations,
a straightforward calculation shows that the additional contribution
to the equations of motion can be represented in terms of the two
scalar functions
\ba
F_{1} (R^2) &=& ({1}/{30})\, 
\left[\,9(\rho')^{4} + 6(\rho'')^{4} + 8\rho'\rho''' - 25\rho''(\rho')^{2} 
-\rho''''\,\right]\ ;\no
F_{2} (R^2) &=& ({1}/{30})\,
\left[\,8 (\rho'')^2 + 4 \rho' \rho''' -
9\rho'' (\rho')^2 - \rho''''\,\right]\ .
\ea
On the one hand, it is apparent that the difference between 
the fifth and any other component of eq.s~\gl{eqgrav3} 
does not include the cosmological constant $\lambda$, since we have
\ba
\rho''  = 
 \frac{{\overline \kappa}}{M^2}
\left\{\Phi^{\prime\,2} + H^{\,\prime\,2} + \frac12
\left(\rho''-\frac13\,{d^2\over dz^2} -\frac13\,\rho'\,{d\over dz}\right)
\left(\Phi^2 + H^2\right) + \frac12\,F_2(R^2)\right\} 
\label{eqgrav051}
\ea
with ${\overline \kappa}$ being defined in eq.~\gl{newcoup}.
On the other hand -- compare with eq.~\gl{cosmcon} -- 
the fifth component actually represents 
the integral of motion for the remaining three equations \gl{eqscalar1} and 
\gl{eqgrav051}, in which $\lambda$ plays the role of an integration constant 
\ba
2 M^2\lambda_{\rm eff} &=&  
\Phi^{\prime\,2} + H^{\,\prime\,2} + 2 \Delta_1\Phi^2 + 2\Delta_2 H^2
- \left(\Phi^2 + H^2\right)^2\no
&-& \frac{4 M^2}{{\overline \kappa}}\,\rho^{\prime\,2} + \left(2\rho^{\prime\,2}  
- \frac43\,\rho'\,{d\over dz}\right)\left(\Phi^2 + H^2\right)\no
&+& F_2(R^2) - F_1(R^2)\ ,
\label{intcosm1}
\ea
where the definition for $\lambda_{\rm eff}$ is adopted from eq.\gl{coseff}.
As a matter of fact, the very last statement can be proved by differentiation 
of the above relationship and substitution of 
eq.s~\gl{eqscalar1} and \gl{eqgrav051}. Moreover, the use of the following identity 
turns out to be  crucial: namely,
\be
4 \rho' F_2(R^2) = {d\over dz}\,\left[\,F_2(R^2) - F_1(R^2)\,\right]\ .
\ee
The latter result is a direct consequence of the general covariance of the
equations of motion.


\section{Gravitational field equation: general metric}

Our starting point is the low energy effective  Action
involving scalar and gravitational fields: namely,
\ba
&&S_{\rm eff} (\Phi, H, g)\ \equiv\
\frac{N\Lambda}{4\pi^3}\,\int d^5X\ \sqrt{g(X)}\ \left\lbrace\,
-\ \frac{2\pi^3}{N\kappa{\cal G}}\,R(X) + \frac{4\pi^3}{N{\kappa\cal G}}\,\lambda\,\right.\no
&&+\ \partial_A \Phi(X)\partial^A \Phi(X)+
\partial_A H(X)\partial^A H(X) - 2 \Delta_1\Phi^2(X)
- 2\Delta_2 H^2(X)\no
&&+\ \left.\left[\,\Phi^2(X) + H^2(X)\,\right]^2 +
\frac16\,R(X)\,\left[\,
\Phi^2(X) + H^2(X)\,\right]\,\right\rbrace
\la{loweff}
\ea
The first variation
$\delta g^{AB}$ with respect to the metric leads to the equations
of motion for the gravitational field.
Taking into account the identities (see \cite{wald}, p.453)
\ba
\delta\sqrt{g} &=& -\ \frac12\,\sqrt{g}\,g_{AB}\,\delta g^{AB}\ ,\\
g^{AB}\,\delta R_{AB} &=& -\ D_A\left(D_B\,\delta g^{AB}\ +\
g^{CD}\, D^A\, \delta g_{CD}\right)\ ,
\ea
we find
\ba
\delta S_{\rm eff}[\,\Phi,H,g\,] =
&-& \frac{N\Lambda}{8\pi^3}\int d^5 X \sqrt{g}
\left\{\,\partial_C \Phi(X)\partial^{\,C} \Phi(X)+
\partial_C H(X)\partial^C H(X)\right.\no
&-&\left. 2\Delta_1 \Phi^2(X) - 2\Delta_2 H^2(X) + [\,\Phi^2(X) + H^2(X)\,]^2\right.\no
&+&\left. {4\pi^3\lambda\over N\kappa{\cal G}}
+\frac{R(X)}{6}\left[\,
\Phi^2(X) + H^2(X) - \frac{12\pi^3}{N\kappa{\cal G}}\,\right]\,\right\} g_{AB}\,\delta g^{AB}\no
&+&\frac{N\Lambda}{4\pi^3}\int d^5 X\sqrt{g}\
\left\{\,\partial_A \Phi(X)\partial_B \Phi(X)\, +
\partial_A H(X)\partial_B H(X)\right.
\no
&+&\left.\frac16 \
R_{AB}(X)\left[\,\Phi^2(X) + H^2(X) 
- {12\pi^3}/{N\kappa{\cal G}}\,\right]\right\}\delta g^{AB}\no
&-&\frac{N\Lambda}{24\pi^3}\int d^5 X\sqrt{g}\,
\left[\,\Phi^2(X) + H^2(X) + {12\pi^3}/{N\kappa{\cal G}}\,\right]\no
&\times&  D_A\left(D_B\,\delta g^{AB}\ +\
g^{CD}\, D^A\, \delta g_{CD}\right)\ .
\ea
With the help of identity\quad
$\Gamma_{AB}^A = \partial_A\ln\sqrt{g}$\quad
the very last integral can be calculated by parts yielding
\ba
&&\frac{N\Lambda}{24\pi^3}\int d^5 X\sqrt{g}\,
 \left[\,
\Phi^2(X) + H^2(X) - {12\pi^3}/{N\kappa{\cal G}}\,\right]
D_A\left(D_B\,\delta g^{AB}\ +\
g^{CD}\, D^A\, \delta g_{CD}\right)\no
&&=\frac{N\Lambda}{24\pi^3}\int d^5 X\sqrt{g}\,
\left\{ D_B\,\partial_A\left[\,\Phi^2(X) + H^2(X)\,\right]
-\ g_{AB}\,D^2\left[\,\Phi^2(X) + H^2(X)\,\right]\right\}
\delta g^{AB}\no
&&-
\frac{N\Lambda}{24\pi^3}\int d^5 X\,\partial_A{\cal B}^A(X)\ ,
\ea
where the boundary term takes the form
\ba
{\cal B}^A(X) &\equiv&
\sqrt{g}\left[\,
\Phi^2(X) + H^2(X) - {12\pi^3}/{N\kappa{\cal G}}\,\right]\no
&\times&\left[\,D_B\,\delta g^{AB}(X)\ -\
g_{CD}(X)\,g^{AB}(X)\,D_B\, \delta g^{CD}(X)\,\right]\no
&-&
\sqrt{g}\left[\,\delta g^{AB}(X) - g^{AB}(X)\,
g_{CD}(X)\,\delta g^{CD}(X)\,\right]\no
&\times&\partial_B\left[\,\Phi^2(X) + H^2(X)\,\right]\ .
\ea
so that we finally obtain, in the Euclidean case and up to the boundary term, the Einstein's
equation in the presence of the cosmological term: namely,
\be
R_{AB} -\frac12\,g_{AB}\left(R - 2\lambda\right)\ =\
\frac{N\kappa{\cal G}}{2\pi^3}\,t_{AB}
\ee
where the normalized energy--momentum tensor of the scalar matter reads
\ba
t_{AB}\ &\equiv&\
\partial_A \Phi\,\partial_B\Phi +
\partial_A H\,\partial_B H\no
&-&\ \frac12\,g_{AB}\left[\,
\partial_C \Phi\, \partial^C \Phi +
\partial_C H\,\partial^C H - 2 \Delta_1\Phi^2
- 2\Delta_2 H^2 + \left(\Phi^2 + H^2\right)^2\,\right]\no
&+&\  \frac16\left(R_{AB} - \frac12\,g_{AB}\,R
+ g_{AB}\,D^C\partial_C  - D_B\,\partial_A\right)
\left(\Phi^2 + H^2\right)\ .
\ea

\end{document}